\documentclass[lettersize,journal]{IEEEtran}
\usepackage{cite}
\usepackage{amsmath,amssymb,amsfonts,amsthm,mathrsfs}
\usepackage{newtxmath}
\usepackage{algorithm,algorithmic}
\usepackage{array}
\usepackage{caption}
\usepackage{textcomp}
\usepackage{stfloats}
\usepackage[hyphens]{url}
\usepackage{verbatim}
\usepackage{graphicx}
\usepackage{subcaption}
\usepackage{booktabs}
\usepackage{xcolor}
\usepackage{soul}
\usepackage{lettrine}
\usepackage{orcidlink}
\usepackage{derivative}
\usepackage{breqn}
\def\BibTeX{{\rm B\kern-.05em{\sc i\kern-.025em b}\kern-.08em
    T\kern-.1667em\lower.7ex\hbox{E}\kern-.125emX}}

\theoremstyle{definition}
\newtheorem{theorem}{Theorem}
\newtheorem{lemma}{Lemma} 

\newcommand{\hermitian}{\text{H}}
\newcommand{\transpose}{\text{T}}
\newcommand{\degrees}{$^{\circ}$}
\newcommand{\vardegrees}{^{\circ}}

\newcommand\numberthis{\addtocounter{equation}{1}\tag{\theequation}}

\usepackage{balance}
\begin{document}

\title{Low-Complexity Optimization of Antenna Switching Schemes for Dynamic Channel Sounding
\thanks{The authors are with the Department of Electrical and Information Technology, Lund University, Lund, Sweden (e-mail: juan.sanchez@eit.lth.se; xuesong.cai@eit.lth.se, ali.al-ameri@eit.lth.se; fredrik.tufvesson@eit.lth.se).

X. Cai is also with Peking University, 100871 Beijing, China (e-mail: xuesong.cai@pku.edu.cn).
}
}

\author{
\IEEEauthorblockN{Juan Sanchez\orcidlink{0000-0001-8392-1697}, \textit{Student Member, IEEE}, Xuesong Cai\orcidlink{0000-0001-7759-7448}, \textit{Senior Member, IEEE},\\ Ali Al-Ameri\orcidlink{0009-0000-8399-9929}, \textit{Student Member, IEEE}, and Fredrik Tufvesson\orcidlink{0000-0003-1072-0784}, \textit{Fellow, IEEE}}

}

\maketitle

\begin{abstract}
Understanding wireless channels is crucial for the design of wireless systems. For mobile communication, sounders and antenna arrays with short measurement times are required to simultaneously capture the dynamic and spatial channel characteristics. Switched antenna arrays are an attractive option that can overcome the high cost of real arrays and the long measurement times of virtual arrays. Optimization of the switching sequences is then essential to avoid aliasing and increase the accuracy of channel parameter estimates. This paper provides a novel and comprehensive analysis of the design of switching sequences. We first review the conventional spatio-temporal ambiguity function, extend it to dual-polarized antenna arrays, and analyze its prohibitive complexity when applied to ultra-massive antenna arrays. We thus propose a new method that uses the Fisher information matrix to tackle the estimation accuracy. We also propose to minimize the ambiguity by choosing a switching sequence that minimizes side lobes in its Fourier spectrum. In this sense, we divide the sequence design problem into Fourier-based ambiguity reduction and Fisher-based accuracy improvement, and coin the resulting design approach as Fourier-Fisher. Simulations and measurements show that the Fourier-Fisher approach achieves identical performance and significantly lower computational complexity than that of the conventional ambiguity-based approach.

\end{abstract}

\section{Introduction} \label{sec:introduction}

\lettrine{U}{nderstanding} wireless propagation channels is a fundamental prerequisite for the design, optimization and verification of wireless systems. As wireless systems evolve, the channel models become increasingly refined in domains like delays, angles, Doppler frequency, birth-and-death behavior of multipath components (MPC), etc. \cite{andrews2014will,wang20206g,cai2024toward,cai2024switched}. Measurement-based channel characterization is irreplaceable, especially for millimeter-wave (mmWave) and sub-THz massive multiple-input multiple-output (MIMO) communications towards 6G, where simulations such as ray tracing can be overly complicated and unrealistic. To measure the spatial characteristics of dynamic channels, channel sounders using antenna arrays with short measurement times are required. There are mainly three different array types, i.e., virtual, real and switched antenna arrays (more illustrative details in \cite{molisch2010wireless}). Virtual arrays consist of moving antennas under different paradigms. The traditional mechanically moving antennas are easy to realize, but are rather slow and almost only applicable to static scenarios. The more general variations, referred to as movable antennas \cite{zhu2025tutorial}, either continue being slow or they cannot realize dynamic massive MIMO sounders. Real arrays consist of multiple radio frequency (RF) chains connected to multiple antennas. They allow for very fast sounding but have issues with complexity, calibration, cost, data storage, etc. Switched arrays consist of a single RF chain connected to different antenna elements that are activated each at a time through an RF switch matrix. Switched arrays strike a balance between virtual and real arrays, becoming an attractive option for dynamic double-directional sounding at high frequencies.


\begin{figure}
     \centering
     \begin{subfigure}[b]{0.48\textwidth}
         \centering
         \includegraphics[width=\textwidth]{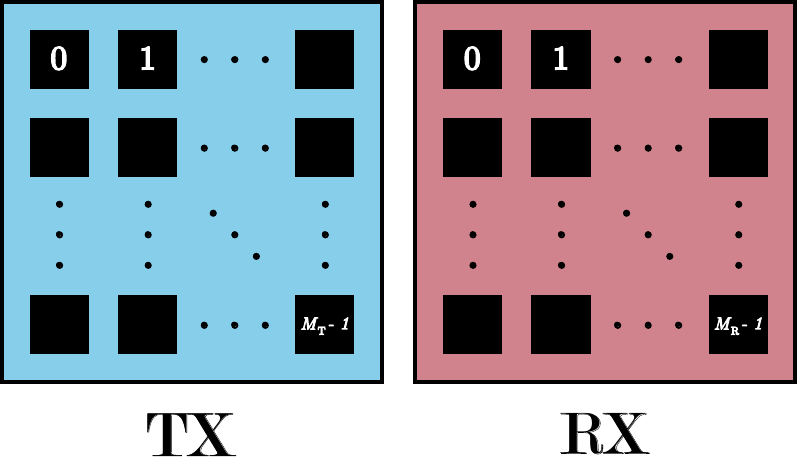}
         \caption{Physical layout}
         \label{fig:trivialSwitchingLayout}
     \end{subfigure}
     \hfill
     \begin{subfigure}[b]{0.48\textwidth}
         \centering
         \includegraphics[width=\textwidth]{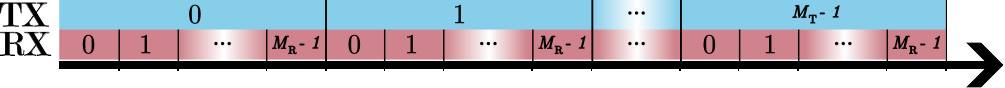}
         \caption{Activation timelines}
         \label{fig:trivialSwitchingTimelines}
     \end{subfigure}
     \caption{An example of trivial (sequential) switching for a MIMO scenario.} \label{fig:trivialSwitching}
\end{figure}

A key consideration in switched arrays is the activation order of the antenna elements. The most trivial switching approach activates the physically adjacent antenna elements in sequential order, as indicated in Fig. \ref{fig:trivialSwitching}. However, this leads to increased ambiguity and constrains the observable Doppler range to the inverse of the snapshot time. In other words, it cannot measure dynamic channels with high-speed scatterers or transceivers. To cover dynamic environments, it is essential to design a sequence that extends the Doppler range to the inverse of the switching rate. The studies in \cite{yin2003doppler,pedersen2004joint} discussed the limitations when using trivial switching sequences in more detail. For a system with single-polarized isotropic antennas, the authors also showed the potential of optimizing the switching sequences to improve the non-ambiguous range in which parametric estimation algorithms such as maximum likelihood estimation (MLE) and its variants can work. However, these studies focus on theoretical constructions of isotropic antennas without considering realistic antenna arrays. Several studies have continued working on switching sequence design with different focus. In \cite{pedersen2008optimization} and \cite{avital2021design}, the sequence design was approached by analyzing the Fisher information matrix (FIM), still under the assumption of arrays of isotropic antenna elements. The authors in \cite{pedersen2008optimization} considered a so-called aperture spatio-temporal matrix to characterize sounders working under different switching sequences. They also provided valuable insights on the effect of the switching configuration on the FIM and the estimation error. Furthermore, the authors introduced a spatio-temporal ambiguity function to characterize ambiguity in the parametric estimation. This function is a direct result of influences in the radar system field, where ``the ambiguity function is a standard means to assess the resolution ability of radar waveforms''. The readers are referred to \cite{woodward2014probability,rendas1998ambiguity,van2004detection,schmidt1986multiple, eric1998ambiguity} for further information on the ambiguity function. The authors in \cite{avital2021design} briefly touched upon the connection between MLE and the Fourier transform of a received signal, analyzed the periodicity of the estimation precision coming from the FIM derivations, and considered switching delays observed in real sounder implementations for the overall switching sequence design. The authors in \cite{wang2019channel} incorporated realistic arrays into the spatio-temporal ambiguity function and proposed a simulated annealing algorithm to solve the sequence design problem.

However, the above studies only considered single-polarized antenna arrays, which excludes the applicability of the theory to dual-polarized antenna arrays. To the best knowledge of the authors, there are no studies that extend the switching sequence design theory to dynamic double-directional dual-polarized switched-array channel sounders. Moreover, the computation of the different ambiguity function variants for single-polarized spatio-temporal arrays is complex and unfeasible for a massive number of antennas, as is the case in modern radio systems. The theory and solutions to the switching problem thus also require a new perspective that can reconcile previous outcomes and provide more efficient switching design solutions.

The present work is not only building on top of the previous results on switching sequences, but also reformulating the design problem into a novel method that emerges from the parametric estimation theory. The proposed approach works for polarimetric channel sounders and can be easily extended to wideband scenarios. The main contributions in this work are:
\begin{itemize}
    \item We develop a \textit{polarimetric} spatio-temporal ambiguity function that incorporates realistic arrays, and analyze the computational gain when the arrays have a high cross-polarization ratio (XPR);
    \item We derive an \textit{estimation accuracy} optimization function based on the FIM of the channel parameters;
    \item We derive an \textit{ambiguity} minimization function based on the Fourier transform of switching sequences;
    \item We solve the sequence design problem for realistic arrays using ambiguity and estimation accuracy steps, referring to the encompassing solution as of \textit{Fourier-Fisher} (FF) approach.
    \item We compare the performance and computational complexity of both ambiguity function and FF approaches via simulations and measurements.
\end{itemize}
The FF approach enables the use of realistic arrays and is shown to require a significantly lower computation time to optimize a switching sequence, compared to the ambiguity approach in \cite{wang2019channel}. This efficiency renders the FF approach suitable for sequence design in ultra-massive MIMO arrays and thus relevant for the future of channel sounding. In addition, covariances between the information scores of different channel parameters (and thus error propagation among their estimates) are reduced more efficiently for realistic antenna arrays with the FF approach, since they are directly addressed when minimizing the off-diagonal elements of the FIM. This has the potential to perform better at high signal-to-noise ratio (SNR) regions, where the multidimensional width of the main estimation peak is of interest.

The remainder of this paper is structured as follows. Sect.\,\ref{sec:model} establishes the general signal model of dual-polarized dynamic MIMO channel sounding. Sect.\,\ref{sec:polarimetricAmbiguity} introduces a polarimetric extension of the spatio-temporal ambiguity function and provides a simplified expression in the case of high XPR. Sect.\,\ref{sec:fourierFisherTheory} focuses on a single polarization pair and develops the theory behind FF switching sequences. Sect.\,\ref{sec:simulationResults} shows realistic simulation results of parametric estimation with MLE using switching sequences previously proposed in the literature and FF switching sequences. Sect.\,\ref{sec:measurementResults} validates the proposed approach with measurements using a mmWave channel sounder. Finally, conclusive remarks are included in Sect.\,\ref{sec:conclusions}.

The notation throughout this paper is as follows. Bold upper case letters, e.g., $\mathbf{B}$, denote matrices. Bold lower case letters, e.g., $\mathbf{b}$, denote column vectors. $\mathbf{B}_i$ denotes the $i$-th column of the matrix $\mathbf{B}$, $[\mathbf{B}]_{ij}$ denotes the element in the $i$-th row and $j$-th column of the matrix $\mathbf{B}$, and $[\mathbf{b}]_{i}$ denotes the $i$-th element of the vector $\mathbf{b}$. $\overline{\mathbf{b}}$ denotes the conjugate of the vector $\mathbf{b}$. The notation $e^{\mathbf{B}}$ expresses the entry-wise exponential function of the matrix $\mathbf{B}$. The superscripts $^{\transpose}$ and $^{\hermitian}$ denote the transpose and the Hermitian transpose, respectively. The operators $\otimes$ and $\odot$ denote the Kronecker and Hadamard products, respectively. The operators $|\mathbf{\cdot}|$ and $||\mathbf{\cdot}||$ denote the absolute value norm and the Euclidean norm, respectively. $\mathbf{I}_n$ denotes the $n \times n$ identity matrix, whereas $\mathbf{1}_n$ denotes the 1-vector of size $n$. For the Fisher information matrix, the notation $[\mathbf{F}]_{\alpha\beta}$ additionally denotes the entry corresponding to the row of the parameter $\alpha$ and the column of the parameter $\beta$.

\section{Signal model}
\label{sec:model}

Consider a switched array channel sounder with $M_{\text{T}}$ antennas at the transmitter (TX) and $M_{\text{R}}$ antennas at the receiver (RX), with centered indexing vectors $\mathbf{m}_{\text{T}} = \left[ 0, \dots, M_{\text{T}} - 1 \right] - \frac{M_{\text{T}} - 1}{2}$ and $\mathbf{m}_{\text{R}} = \left[ 0, \dots, M_{\text{R}} - 1 \right] - \frac{M_{\text{R}} - 1}{2}$, respectively. There are a total of $M_{\text{TR}} = M_{\text{T}} \cdot M_{\text{R}}$ combinations of antenna pairs to be measured per MIMO snapshot\footnote{A snapshot contains the channels of all antenna combinations.}, with centered indexing vector $\mathbf{m}_{\text{TR}} = \left[ 0, \dots, M_{\text{TR}} - 1 \right] - \frac{M_{\text{TR}} - 1}{2}$. $M_t$ MIMO snapshots are taken. We assume that the measurement time of $M_t$ MIMO snapshots is smaller than the coherence time of the channel\footnote{The maximum number of snapshots $M_t$ is determined by the coherence time of the channel. During the coherence time, the structural parameters (angles, Doppler frequencies, delays, etc.) of MPCs are assumed to be unchanged.}, and that the antenna array responses are flat within the measurement bandwidth with $M_f$ frequency points. The general vectorized data model for $P$ MPCs is given by \cite{richter2005estimation}
\begin{equation} \label{eq:general_vectorized_data_model}
  \begin{aligned}
    \mathbf{s}(\boldsymbol{\theta}_{\text{sp}})=  
    \sum_{p=1}^{P} \mathbf{B}(\boldsymbol{\mu}_{p})\cdot\boldsymbol{\gamma}_p,
 \end{aligned}
\end{equation}
where $\boldsymbol{\mu}_p$ includes the structural parameters for the $p$-th path, $\boldsymbol{\gamma}_p \in \mathbb{C}^{4}$ contains the polarimetric transmission coefficients for the $p$-th path, $\boldsymbol{\theta}_{\text{sp}} = \{ \boldsymbol{\mu}_p, \boldsymbol{\gamma}_p: \, p = 1,\dots,P \}$, and $\mathbf{B}(\boldsymbol{\mu}_p) \in \mathbb{C}^{M_t M_{\text{TR}} M_f \times 4}$ is the basis matrix for a single path. For dynamic channels, the basis matrix can be expressed as
\begin{equation} \label{eq:generic_basis_matrix}
 \begin{aligned}
    \mathbf{B}(\boldsymbol{\mu}_{p}) =
        {\begin{bmatrix}
        (((\bold{b}_t \otimes \bold{b}_{\text{T}_{\text{H}}} \otimes \bold{b}_{\text{R}_{\text{H}}}) \odot \bold{a}_{\nu}) \otimes \bold{b}_f)^{\text{T}}\\
        (((\bold{b}_t \otimes \bold{b}_{\text{T}_{\text{H}}} \otimes \bold{b}_{\text{R}_{\text{V}}}) \odot \bold{a}_{\nu}) \otimes \bold{b}_f)^{\text{T}}\\
        (((\bold{b}_t \otimes \bold{b}_{\text{T}_{\text{V}}} \otimes \bold{b}_{\text{R}_{\text{H}}}) \odot \bold{a}_{\nu}) \otimes \bold{b}_f)^{\text{T}}\\
        (((\bold{b}_t \otimes \bold{b}_{\text{T}_{\text{V}}} \otimes \bold{b}_{\text{R}_{\text{V}}}) \odot \bold{a}_{\nu}) \otimes \bold{b}_f)^{\text{T}}\\
        \end{bmatrix}}^{\text{T}},
 \end{aligned}
\end{equation}
where $\bold{b}_t\in \mathbb{C}^{M_t}$ is the Doppler-induced change in responses due to the different starting time instants of the MIMO snapshots, $\bold{b}_{\text{T}_{\text{H}}}, \bold{b}_{{\text{T}_{\text{V}}}}\in \mathbb{C}^{M_{\text{T}}}$ represent the polarimetric TX array responses at the horizontal and the vertical polarizations, respectively,  $\bold{b}_{{\text{R}_{\text{H}}}}, \bold{b}_{{\text{R}_{\text{V}}}}\in \mathbb{C}^{M_{\text{R}}}$ represent the polarimetric RX array responses, and $\bold{b}_f\in \mathbb{C}^{M_f}$ is the frequency basis vector dependent on the path delay. $\bold{b}_{\text{T}_{\text{H}}}, \bold{b}_{{\text{T}_{\text{V}}}}$ are dependent on the elevation of departure (EOD) $\vartheta_{\text{T}}$ and azimuth of departure (AOD) $\varphi_{\text{T}}$. $\bold{b}_{\text{R}_{\text{H}}}, \bold{b}_{{\text{R}_{\text{V}}}}$ are dependent on the elevation of arrival (EOA) $\vartheta_{\text{R}}$ and azimuth of arrival (AOA) $\varphi_{\text{R}}$. Lastly, $\bold{a}_{\nu}\in \mathbb{C}^{M_tM_{\text{TR}}}$ represents the Doppler-induced change of responses for different antenna pairs in every MIMO snapshot. Specifically, 
\begin{equation} \label{eq:phaseVectorDoppler}
\begin{aligned}
    [&\bold{a}_{\nu}]_{m_{\text{TR}}+(m_t-1)\cdot M_{\text{TR}}} = e^{j2\pi{\nu}_p[\boldsymbol{\eta}]_{m_{\text{TR}}+(m_t-1)\cdot M_{\text{TR}}}}, \  \\ & m_{\text{TR}} = 1, \dots, M_{\text{TR}}; \quad m_t = 1, \dots, M_{\text{t}},
\end{aligned}
\end{equation} where $\nu_p$ is the Doppler frequency of the $p$-th MPC, and $[\boldsymbol{\eta}]_{m_{\text{TR}}+(m_t-1)\cdot M_{\text{TR}}}$ is the time instant when the $m_{\text{TR}}$-th antenna pair is activated in the $m_t$-th snapshot relative to the starting time instant of the $m_t$-th snapshot. Note that the activation time instants can be independent of the snapshot index $m_t$, i.e., $[\boldsymbol{\eta}]_{m_{\text{TR}}+(m_t-1)\cdot M_{\text{TR}}} = [\boldsymbol{\eta}]_{m_{\text{TR}}+(m'_t-1)\cdot M_{\text{TR}}}$, $\forall m'_t \neq m_t$. The structure of the vector $\boldsymbol{\eta}$ for a single snapshot can be represented as
\begin{equation} \label{eq:switchingVectorCanonical}
    \boldsymbol{\eta} = \mathbf{m}_{\text{TR}} \mathbf{P}_\pi \cdot \Delta t,
\end{equation}
where $\mathbf{P}_\pi \in \mathbb{N}^{M_{\text{TR}} \times M_{\text{TR}}}$ is any permutation matrix, and $\Delta t$ is the time difference between activating two different antenna pairs. An alternative representation is $\boldsymbol{\eta} \in \mathfrak{N}_{M_{\text{TR}}}$, where $\mathfrak{N}_{M_{\text{TR}}} = \left\{ \Delta t \cdot \left( \mathbf{x} - 1 - \frac{M_{\text{TR}} - 1}{2} \right): \mathbf{x} \in \mathfrak{G}_{M_{\text{TR}}} \right\}$. The set $\mathfrak{G}_{M_{\text{TR}}}$ represents the symmetric group on the set $\{ 1, 2, \dots, M_{\text{TR}} \}$, i.e. a group that consists of all the permutations that can be performed on the set $\{ 1, 2, \dots, M_{\text{TR}} \}$. More details of the data model can be found in \cite{wang2017high}, \cite{mota2010estimation}.

\section{Polarimetric spatio-temporal ambiguity function}
\label{sec:polarimetricAmbiguity}

Ambiguity has been a problem discussed in the literature since decades ago, with a clear description in \cite{schmidt1986multiple}. It occurs when received signals can be characterized by non-unique combinations of parameters (e.g. angle, Doppler frequency, etc.). The accuracy of parameter estimation is thus worsened by the high likelihood of the combinations that do not correspond to the ground truth. Ambiguity was quantified in \cite{eric1998ambiguity} by evaluating the correlation level between resulting signals of the ground truth parameters and any other parameter combination. The resulting function is called the Ambiguity I function. The spatio-temporal ambiguity function introduced in \cite{wang2019channel} is an extension of the Ambiguity I function that caters to structural parameters towards a double-directional description of the wireless channel. This variant can characterize the performance of different switching sequences $\boldsymbol{\eta}$ when used in switched array channel sounding, and makes use of the enhanced aperture distribution function (EADF) \cite{landmann2004efficient} to handle real-world antenna array responses. The spatio-temporal ambiguity function has the form
\begin{equation}
    \label{eq:ambiguityCanonical}
    \begin{aligned}
    X(\boldsymbol{\mu}_{p},\boldsymbol{\mu}'_{p}; \boldsymbol{\eta}) & = \frac
    {\mathbf{b}^{\text{H}}(\boldsymbol{\mu}_{p}, \boldsymbol{\eta})\mathbf{b}(\boldsymbol{\mu}'_{p}, \boldsymbol{\eta})}
    {||\mathbf{b}^{\text{H}}(\boldsymbol{\mu}_{p}, \boldsymbol{\eta})|| \cdot ||\mathbf{b}(\boldsymbol{\mu}'_{p}, \boldsymbol{\eta})||},
    \end{aligned}
\end{equation}
where the vector $\mathbf{b}$ is the simplification of the general basis matrix $\mathbf{B}$ from (\ref{eq:generic_basis_matrix}) when considering a single polarization pair, and $\boldsymbol{\mu}'_{p}$ is any collection of structural path parameters that includes the true $\boldsymbol{\mu}_{p}$. The ambiguity function is a similarity measure between phase vectors with its modulus restricted to the interval $0 \leq |X(\boldsymbol{\mu}_{p},\boldsymbol{\mu}'_{p})| \leq 1$. Notice the highlighted dependence on $\boldsymbol{\eta}$, which is the essence of the optimization methods aiming to minimize the ambiguity of parametric estimation and extend the estimation range as a consequence. Fig. \ref{fig:ambFunctions} shows examples of ambiguity functions for different switching sequences, for a fixed true azimuth $\varphi = 90\vardegrees$, azimuth search space $\varphi' \in [-180,180]\vardegrees$, and variable Doppler difference $\Delta \nu = \nu - \nu'$, where $\nu$ is the true Doppler and $\nu'$ is any possible Doppler within the search space. In this context, a single path is considered, and only azimuth of departure and Doppler are unknown, which implies that $\boldsymbol{\mu}_p=(\varphi,\nu)$ and $\boldsymbol{\mu}'_p=(\varphi',\nu')$.

The authors in \cite[(20)]{wang2019channel} formulate an optimization problem to design switching sequences for a linear array by considering only the azimuth angle on one side and leaving the switching sequence on the other side unoptimized. The authors in \cite{al2023hybrid} further extended the formulation with an objective function (to be minimized) that considers the azimuths and elevations on both the TX and RX sides:
\begin{equation} \label{eq:objectiveAmbiguity}
    f_{\mathcal{P}}(\boldsymbol{\eta}) =  \iint_D |X({\boldsymbol{\mu}_p, \boldsymbol{\mu}'_p, \boldsymbol{\eta})}|^{\mathcal{P}} \,d\boldsymbol{\mu}_p\,d\boldsymbol{\mu}'_p,
\end{equation}
where $\mathcal{P}$ is a tuning parameter, $\boldsymbol{\mu}_p=(\varphi_{\text{T}},\vartheta_{\text{T}},\varphi_{\text{R}},\vartheta_{\text{R}},\nu_p)$ and $\boldsymbol{\mu}'_p=(\varphi_{\text{T}}',\vartheta_{\text{T}}',\varphi_{\text{R}}',\vartheta_{\text{R}}',\nu'_p)$ are the true and test structural parameters of the channel for the $p$-th path, 
and the integration region $D$ is defined as
\begin{equation} \label{}
  \begin{aligned}
        D= \{\varphi_{\text{T}},\varphi_{\text{T}}', \varphi_{\text{R}},\varphi_{\text{R}}' \in[0,&\ 2\pi] \ \&   \\ \ \vartheta_{\text{T}}, \vartheta_{\text{T}}', \vartheta_{\text{R}}, \vartheta_{\text{R}}'\in[0,\ \pi]\ & \& \ \nu_p-\nu'_p\in[-\nu_{\text{up},p},\nu_{\text{up},p}] \}.
   \end{aligned}
\end{equation}
The objective function esentially integrates over the entire set of possible true $\boldsymbol{\mu}_p$ and test $\boldsymbol{\mu}'_p$ structural parameter combinations, and penalizes multiple peaks that are found in its image. As an optimal switching sequence is to be found, the minimization problem is expressed as $\min_{\boldsymbol{\eta}} f_{\mathcal{P}}(\boldsymbol{\eta})$.

Since (\ref{eq:ambiguityCanonical}) only considers a single polarization pair, it fails to characterize the overall ambiguity present in all the four polarization pairs of a polarimetric signal model. The switching design from (\ref{eq:objectiveAmbiguity}) is also affected by the choice of the tuning parameter $\mathcal{P}$ utilized within the optimization process. For a generalized polarimetric switching sequence design, it is necessary to establish a measure between the basis matrices as the ambiguity function for the optimization process.

\subsection{General case}

The theory behind canonical angles allows us to calculate distances between subspaces, which in turn can be characterized by the column space of a matrix. This means that it is possible to formulate a polarimetric ambiguity measure by computing a similarity measure between the subspaces spanned by the columns of the basis matrices to be compared. Mathematically,
\begin{equation} \label{eq:ambiguitySubspaces}
    X(\boldsymbol{\mu}_p,\boldsymbol{\mu}_p') = f_{\text{Dist}}(\mathcal{B}, \mathcal{B}'),
\end{equation}
where $\mathcal{B} = \text{span}\{\mathbf{B}_i(\boldsymbol{\mu}_{p}), \, i=1, \dots, 4\}, \mathcal{B}' = \text{span}\{\mathbf{B}_i(\boldsymbol{\mu}_{p}'), \, i=1, \dots, 4\}$. Using QR decomposition, a basis matrix $\mathbf{B}$ can be decomposed into $\mathbf{B} = \mathbf{Q} \mathbf{R}$, where $\mathbf{Q} \in \mathbb{C}^{ M_{\text{TR}} \times M_{\text{TR}} }$ is unitary, and $\mathbf{R} \in \mathbb{C}^{ M_{\text{TR}} \times 4 }$ is in echelon form. Note from (\ref{eq:generic_basis_matrix}) that $\mathbf{B}$ contains the products of polarimetric responses $\mathbf{b}_{\text{T}_{\text{H}}} \otimes \mathbf{b}_{\text{R}_{\text{H}}}$, $\mathbf{b}_{\text{T}_{\text{H}}} \otimes \mathbf{b}_{\text{R}_{\text{V}}}$, $\mathbf{b}_{\text{T}_{\text{V}}} \otimes \mathbf{b}_{\text{R}_{\text{H}}}$, $\mathbf{b}_{\text{T}_{\text{V}}} \otimes \mathbf{b}_{\text{R}_{\text{V}}}$ within its columns. The polarimetric responses do not exhibit perfect correlation (but rather weak correlation) since the cross-polarization radiation pattern of an antenna is practically of much lower intensity and non-symmetrical in comparison with its co-polarized pattern \cite{balanis2016antenna}. Given that perfect correlation between vectors is a sufficient and necessary condition for linear dependence, it follows that the column vectors $\mathbf{B}_i$ are linearly independent. Thus, $\text{rank}(\mathbf{B})$ = 4 and the first four columns of $\mathbf{Q}$ form an orthonormal basis for the column space of $\mathbf{B}$. This implies that $\text{span}\{\mathbf{Q}_i(\boldsymbol{\mu}_{p})\} = \text{span}\{\mathbf{B}_i(\boldsymbol{\mu}_{p})\} = \mathcal{B}, i = 1, \dots, 4$. Let us include these columns into the matrix $\mathbf{Q}_{\mathcal{B}} = \left[ \mathbf{Q}_1 \mathbf{Q}_2 \mathbf{Q}_3 \mathbf{Q}_4 \right]$. Now $\mathbf{Q}_{\mathcal{B}}$ meets the sufficient condition to use the Grassmanian distance as the distance between subspaces. Normalizing to the rank, (\ref{eq:ambiguitySubspaces}) can be then further developed into
\begin{equation} \label{eq:distanceMatrices}
    d_2(\mathcal{B}, \mathcal{B}') = \frac{1}{4} \cdot \left( \sum_{i=1}^4 (\arccos{\mathbf{\Sigma}_{ii}})^2 \right)^{1/2},
\end{equation}
where $\mathbf{\Sigma}$ denotes the matrix containing the singular values of the SVD decomposition $\mathbf{Q}_{\mathcal{B}\mathcal{B}'} = \mathbf{Q}_{\mathcal{B}}^{\text{H}}(\boldsymbol{\mu}_{p})\mathbf{Q}_{\mathcal{B}'}(\boldsymbol{\mu}_{p}') = \mathbf{U}\mathbf{\Sigma}\mathbf{V}^{\text{H}}$. (\ref{eq:distanceMatrices}) can be seen as the 2-norm of the vector of principal angles of the multiplication matrix. Since the principal angles are the arccosines of the singular values of the multiplication matrix, a similarity measure can be realized by looking at the singular values, still using the 2-norm. Then, the measure has the form
\begin{equation} \label{eq:proxMatrices2Norm}
    S_2(\boldsymbol{\mu}_p,\boldsymbol{\mu}_p') = \frac{1}{4} \cdot \left( \sum_{i=1}^4 (\mathbf{\Sigma}_{ii})^2 \right)^{1/2}.
\end{equation}
Since all norms are equivalent \cite{horn2012matrix}, the 1-norm can also act as the base for a more efficient and final similarity measure form. The resulting similarity measure using the 1-norm, i.e. the polarimetric spatio-temporal ambiguity function, based on the Grassmanian distance can be expressed as
\begin{equation} \label{eq:ambiguityMatrices}
    X(\boldsymbol{\mu}_p,\boldsymbol{\mu}_p') = f_{\text{Dist}}(\mathcal{B}, \mathcal{B}') = S_1(\boldsymbol{\mu}_p,\boldsymbol{\mu}_p') = \frac{1}{4} \cdot \sum_{i=1}^4 \mathbf{\Sigma}_{ii}.
\end{equation}
Notice that the image of the ambiguity function is $[0,1]$, since $\mathbf{\Sigma}_{ii} \in [0,1]$, $\forall i=1, \dots, 4$.\\

\subsection{High cross-polarization ratio case}\label{ss:ambiguityHighXPR}

In the particular case of high XPR (e.g., $\geq$ 30\,dB) at both TX and RX sides, the array responses satisfy
\begin{equation} \label{eq:orthogonalityHighXPR}
\begin{gathered}
    [ \mathbf{b}_{\text{T}_{\text{H}}} ]_i \cdot  [ \mathbf{b}_{\text{T}_{\text{V}}} ]_i \approx 0,\quad [ \mathbf{b}_{\text{T}_{\text{H}}} ]_i +  [ \mathbf{b}_{\text{T}_{\text{V}}} ]_i = c_i,\\
    [ \mathbf{b}_{\text{R}_{\text{H}}} ]_j \cdot  [ \mathbf{b}_{\text{R}_{\text{V}}} ]_j \approx 0,\quad [ \mathbf{b}_{\text{R}_{\text{H}}} ]_j +  [ \mathbf{b}_{\text{R}_{\text{V}}} ]_j = d_j,\\
    \forall i = 1, \dots, M_{\text{T}},\quad j = 1, \dots, M_{\text{R}},
\end{gathered}
\end{equation}
where $c_i$ and $d_j$ are constants. This means that an antenna element is either almost perfectly horizontally polarized or almost perfectly vertically polarized. This specific structure of the array responses allows for a computationally more efficient form of the ambiguity function, whose derivation will be supported by the following theorems.\\

\begin{theorem}
\label{theorem:kroneckerOrthogonality}
The Kronecker product preserves the orthogonality between the vectors. In other words, $\forall \mathbf{u},\mathbf{v},\mathbf{w} \in \mathbb{C}^n$,
\begin{equation*}
    \langle \mathbf{u},\mathbf{v} \rangle = 0 \quad \Rightarrow \quad \langle \mathbf{u} \otimes \mathbf{w} , \mathbf{v} \otimes \mathbf{w} \rangle = \langle \mathbf{w} \otimes \mathbf{u} , \mathbf{w} \otimes \mathbf{v} \rangle = 0.
\end{equation*}
\end{theorem}
\begin{proof}
See Appendix \ref{app:proofKroneckerOrthogonality}.
\end{proof}

\begin{theorem}
\label{theorem:hadamardOrthogonality}
$\forall \mathbf{u},\mathbf{v},\mathbf{w} \in \mathbb{C}^n$, if $u_i \cdot v_i = 0$ then
\begin{equation*}
    \langle \mathbf{u} \circ \mathbf{w} , \mathbf{v} \circ \mathbf{w} \rangle =
    \langle \mathbf{w} \circ \mathbf{u} , \mathbf{w} \circ \mathbf{v} \rangle = 0.
\end{equation*}
\end{theorem}
\begin{proof}
See Appendix \ref{app:proofHadamardOrthogonality}.
\end{proof}

The inner product between $\mathbf{b}_{\text{T}_{\text{H}}}$ and $\mathbf{b}_{\text{T}_{\text{V}}}$, given their structure in (\ref{eq:orthogonalityHighXPR}), can be computed as

\begin{equation}
    \langle \mathbf{b}_{\text{T}_{\text{H}}} , \mathbf{b}_{\text{T}_{\text{V}}} \rangle = \sum_i [ \overline{ \mathbf{b}_{\text{T}_{\text{H}}} ]_i } \cdot  [ \mathbf{b}_{\text{T}_{\text{V}}} ]_i = \sum_i 0 = 0.
\end{equation}
The same result is obtained when computing $\langle \mathbf{b}_{\text{R}_{\text{H}}} , \mathbf{b}_{\text{R}_{\text{V}}} \rangle$. Hence, the vectors $\mathbf{b}_{\text{T}_{\text{H}}}$,  $\mathbf{b}_{\text{T}_{\text{V}}}$, and $\mathbf{b}_{\text{R}_{\text{H}}}$, $\mathbf{b}_{\text{R}_{\text{V}}}$ are pairwise orthogonal. Using Theorems \ref{theorem:kroneckerOrthogonality} and \ref{theorem:hadamardOrthogonality}, if the array responses are defined according to (\ref{eq:orthogonalityHighXPR}), the vectors
\begin{equation*}
    \mathbf{b}_{\text{T}_{\text{H}}} \otimes \mathbf{b}_{\text{R}_{\text{H}}},\quad \mathbf{b}_{\text{T}_{\text{H}}} \otimes \mathbf{b}_{\text{R}_{\text{V}}},\quad
    \mathbf{b}_{\text{T}_{\text{V}}} \otimes \mathbf{b}_{\text{R}_{\text{H}}},\quad \mathbf{b}_{\text{T}_{\text{V}}} \otimes \mathbf{b}_{\text{R}_{\text{V}}},
\end{equation*}
are orthogonal to each other. It is thus clear that the matrix $\mathbf{B}(\boldsymbol{\mu}_p)$ is composed of orthogonal vectors. To have a matrix $\mathbf{Q_{\mathcal{B}}}$ whose columns constitute an orthonormal basis spanning the same subspace as $\mathbf{B}(\boldsymbol{\mu}_p)$, it is sufficient to normalize each vector $\mathbf{B}_i(\boldsymbol{\mu}_{p})$ as $\displaystyle \mathbf{Q}_{\mathcal{B},i} = \frac{\mathbf{B}_i(\boldsymbol{\mu}_{p})}{||\mathbf{B}_i(\boldsymbol{\mu}_{p})||}$. Each element of the matrix $\mathbf{Q}_{\mathcal{B}\mathcal{B}'}$ then follows the structure
\begin{equation} \label{eq:componentsQXPR}
\begin{aligned}
    [\mathbf{Q}_{\mathcal{B}\mathcal{B}'}]_{ij}
    &= \sum_k [\mathbf{Q}_{\mathcal{B}}^{\text{H}}(\boldsymbol{\mu}_{p})]_{ik} [\mathbf{Q}_{\mathcal{B}'}(\boldsymbol{\mu}_{p}')]_{kj}\\
    &= \sum_k \overline{ [\mathbf{Q}_{\mathcal{B}}(\boldsymbol{\mu}_{p})]_{ki} } [\mathbf{Q}_{\mathcal{B}'}(\boldsymbol{\mu}_{p}')]_{kj}.
\end{aligned}
\end{equation}
By using property (\ref{eq:orthogonalityHighXPR}) in (\ref{eq:componentsQXPR}), $\forall i = 1, \dots, 4$,
\begin{equation} \label{eq:diagHighXPR}
\begin{aligned}
    \mathbf{Q}_{\mathcal{B}\mathcal{B}'}
    &= \text{diag} \left\{ \sum_k \overline{ [\mathbf{Q}_{\mathcal{B}}(\boldsymbol{\mu}_{p})]_{ki} } [\mathbf{Q}_{\mathcal{B}'}(\boldsymbol{\mu}_{p}')]_{ki} \right\}\\
    &= \text{diag} \left\{ \frac{\mathbf{B}_i^{\text{H}}(\boldsymbol{\mu}_{p}) \mathbf{B}_i(\boldsymbol{\mu}_{p}')}{||\mathbf{B}_i^{\text{H}}(\boldsymbol{\mu}_{p})|| \cdot ||\mathbf{B}_i(\boldsymbol{\mu}_{p}')||} \right\}.
\end{aligned}
\end{equation}
Since $\mathbf{Q}_{\mathcal{B}\mathcal{B}'}$ is a diagonal matrix, it is possible to find appropriate generalized permutation matrices $U$ and $V$ that exchange the order and sign of the diagonal elements, transforming $\mathbf{Q}_{\mathcal{B}\mathcal{B}'}$ into a diagonal matrix $\Sigma$ of nonnegative entries sorted in descending order. Since a generalized permutation matrix is also a unitary matrix, the matrices $U$, $\Sigma$, and $V$ correspond to the SVD decomposition of $\mathbf{Q}_{\mathcal{B}\mathcal{B}'}$. Since $U$ and $V$ perform operations equivalent to the absolute value and the exchange of the order of diagonal elements, we can perform the absolute value on the elements of $\mathbf{Q}_{\mathcal{B}\mathcal{B}'}$ and use the resulting matrix to calculate the ambiguity function with no effect on the result. Then, (\ref{eq:ambiguityMatrices}) can be written as
\begin{equation} \label{eq:simplifiedAmbiguityXPR}
    X(\boldsymbol{\mu}_p,\boldsymbol{\mu}_p') = \frac{1}{4} \cdot \sum_{i=1}^4 \frac{|\mathbf{B}^{\hermitian}_i(\boldsymbol{\mu}_{p}) \mathbf{B}_i(\boldsymbol{\mu}_{p}')|}{||\mathbf{B}^{\hermitian}_i(\boldsymbol{\mu}_{p})|| \cdot ||\mathbf{B}_i(\boldsymbol{\mu}_{p}')||} = \frac{1}{4} \cdot \sum_{i=1}^4 |X_i|.
\end{equation}
Notice the similarities between (\ref{eq:ambiguityCanonical}) and (\ref{eq:simplifiedAmbiguityXPR}), where the latter can be seen as the normalized sum of the magnitude of the spatio-ambiguity functions $X_i$ associated to each of the polarization pairs.

\subsection{Kronecker switching}\label{ss:kroneckerSwitching}

By expanding the spatio-ambiguity function for a single polarization pair, the conditions under high XPR in (\ref{eq:orthogonalityHighXPR}) can be further exploited. Taking the spatio-ambiguity function for the first column of the basis matrix $\mathbf{B}$ in (\ref{eq:generic_basis_matrix}) results in
\begin{equation}
\begin{aligned}
    X_1(\boldsymbol{\mu}_p,\boldsymbol{\mu}_p') = \frac{\mathbf{B}^{\hermitian}_1(\boldsymbol{\mu}_{p}) \mathbf{B}_1(\boldsymbol{\mu}_{p}')}{||\mathbf{B}^{\hermitian}_1(\boldsymbol{\mu}_{p})|| \cdot ||\mathbf{B}_1(\boldsymbol{\mu}_{p}')||},
\end{aligned}
\end{equation}
with a proportionality relation
\begin{equation} \label{eq:kroneckerProductMiddleStep}
\begin{aligned}
    X_1 \propto& (((\bold{b}_t \otimes \bold{b}_{\text{T}_{\text{H}}} \otimes \bold{b}_{\text{R}_{\text{H}}}) \odot \bold{a}_{\nu}) \otimes \bold{b}_f)^{\hermitian}\\
    & \cdot (((\bold{b}_t' \otimes \bold{b}_{\text{T}_{\text{H}}}' \otimes \bold{b}_{\text{R}_{\text{H}}}') \odot \bold{a}_{\nu}') \otimes \bold{b}_f')\\
    \propto&  (( \bold{b}_{\text{T}_{\text{H}}} \otimes \bold{b}_{\text{R}_{\text{H}}}) \odot \bold{a}_{\nu})^{\hermitian} \cdot (( \bold{b}_{\text{T}_{\text{H}}}' \otimes \bold{b}_{\text{R}_{\text{H}}}') \odot \bold{a}_{\nu}').
\end{aligned}
\end{equation}
Within $\bold{a}_{\nu}$, it is always possible to construct a sequence $\boldsymbol{\eta} = \boldsymbol{\eta}_{\text{T}} \otimes \boldsymbol{\eta}_{\text{R}}$, i.e. the Kronecker product of a TX sequence $\boldsymbol{\eta}_{\text{T}}$ and a RX sequence $\boldsymbol{\eta}_{\text{R}}$. By plugging this construction into (\ref{eq:kroneckerProductMiddleStep}), the proportionality relation becomes
\begin{equation} \label{eq:kroneckerProductAmbiguity}
\begin{aligned}
    X_1 \propto& \left( \left[ (\bold{b}_{\text{T}_{\text{H}}} \odot \boldsymbol{\eta}_{\text{T}})^{\hermitian} \cdot (\bold{b}_{\text{T}_{\text{H}}}' \odot \boldsymbol{\eta}_{\text{T}}) \right] \right.\\
    &\left. \otimes \left[ (\bold{b}_{\text{R}_{\text{H}}} \odot \boldsymbol{\eta}_{\text{R}})^{\hermitian} \cdot (\bold{b}_{\text{R}_{\text{H}}}' \odot \boldsymbol{\eta}_{\text{R}}) \right] \right) \odot \exp(j2 \pi (\nu' - \nu))\\
    \propto& (\bold{b}_{\text{T}_{\text{H}}} \odot \boldsymbol{\eta}_{\text{T}})^{\hermitian} \cdot (\bold{b}_{\text{T}_{\text{H}}}' \odot \boldsymbol{\eta}_{\text{T}}) \cdot (\bold{b}_{\text{R}_{\text{H}}} \odot \boldsymbol{\eta}_{\text{R}})^{\hermitian} \cdot (\bold{b}_{\text{R}_{\text{H}}}' \odot \boldsymbol{\eta}_{\text{R}}),
\end{aligned}
\end{equation}
where the fact that the Kronecker product of two numbers is equal to their product was used. This result implies an independent optimization of the TX and RX sequences and thus a remarkably shorter computation time than that of a joint sequence optimization process.\\

\subsection{Computational complexity of antenna switching design based on the ambiguity function}\label{ss:complexityAmbiguity}

The result in (\ref{eq:simplifiedAmbiguityXPR}) has great implications for the computational complexity of calculating ambiguities for polarimetric channel sounding. If the XPR is high enough, it is possible to separate, and potentially parallelize, the calculation of ambiguities to polarization pairs and later add them up to get the polarimetric spatio-temporal ambiguity function. Using \cite{cormen2022introduction,golub2013matrix} as references, the computational complexities of different numerical operations are collected in Table \ref{tab:complexities}.

\begin{table}
\begin{center}
    \caption{Computational complexity of numerical operations.}
    \label{tab:complexities}
    \begin{tabular}{|c|c|}
	
    \hline		
    \textbf{Operation}	& \textbf{Complexity}\\\hline
    $\mathbb{C}^1$ basic arithmetic & $\mathcal{O}\left(1\right)$ \\\hline
    Vector $l^2$-norm & $\mathcal{O}\left(n\right)$ \\\hline
    Complex transpose & $\mathcal{O}\left(nm\right)$ \\\hline
    Square matrix multiplication & $\mathcal{O}\left(n^3\right)$ \\\hline
    QR decomposition & $\mathcal{O}\left(mn^2\right), m \geq n$ \\\hline
    SVD decomposition & $\mathcal{O}\left(m^2n\right), m \geq n$ \\\hline
    Fast Fourier transform & $\mathcal{O}\left(n \log{n}\right)$ \\\hline
    Median of an $n$-sized set & $\mathcal{O}\left(n\right)$ \\\hline
    
    \end{tabular}
\end{center}
\end{table}

For the general case in (\ref{eq:ambiguityMatrices}), the complexity is $\mathcal{O}\left(2 \cdot 16 M_t M_{\text{TR}} M_f\right) + \mathcal{O}\left(4 \cdot M_t M_{\text{TR}} M_f\right) + \mathcal{O}\left(\left(M_t M_{\text{TR}} M_f\right)^3\right) + \mathcal{O}\left(\left(M_t M_{\text{TR}} M_f\right)^3\right) + \mathcal{O}\left(5\right) = \mathcal{O}\left(\left(M_t M_{\text{TR}} M_f\right)^3\right)$. For the case of high XPR ratio in (\ref{eq:simplifiedAmbiguityXPR}), the complexity is $\mathcal{O}\left(4 \cdot 2 \cdot M_t M_{\text{TR}} M_f\right) + \mathcal{O}\left(4 \cdot M_t M_{\text{TR}} M_f\right) + \mathcal{O}\left(4 \cdot 2 \cdot M_t M_{\text{TR}} M_f\right) + \mathcal{O}\left(4 \cdot 3 + 1\right) = \mathcal{O}\left(M_t M_{\text{TR}} M_f\right)$. This shows the significant complexity reduction when using antenna arrays with high XPR.

The complexity of the objective function in (\ref{eq:objectiveAmbiguity}) becomes $\mathcal{O}\left(M_{\vartheta_{\text{T}}} M_{\varphi_{\text{T}}} M_{\vartheta_{\text{R}}} M_{\varphi_{\text{R}}} M_\nu \mathcal{P} \left( M_t M_{\text{TR}} M_f \right)^3 \right)$ for an ambiguity function under the general case in (\ref{eq:ambiguityMatrices}), and $\mathcal{O}\left(M_{\vartheta_{\text{T}}} M_{\varphi_{\text{T}}} M_{\vartheta_{\text{R}}} M_{\varphi_{\text{R}}} M_\nu \mathcal{P} M_t M_{\text{TR}} M_f\right)$ under the high-XPR case in (\ref{eq:simplifiedAmbiguityXPR}). The variables $M_{{\vartheta}_{\text{T}}}$, $M_{{\varphi}_{\text{T}}}$, $M_{{\vartheta}_{\text{R}}}$, $M_{{\varphi}_{\text{R}}}$ and $M_\nu$ represent the number of discrete points in different directions of arrival/departure and in Doppler frequency used for integration, respectively. A simulated annealing algorithm then uses this function to solve the sequence optimization problem \cite{wang2019channel}. Since simulated annealing is a metaheuristic, its time complexity will be analyzed later in Sect.\,\ref{sec:simulationResults}. However, it is clear from the objective function's complexity that a single iteration in the simulated annealing procedure can take up a prohibitive amount of time.\\

\section{Antenna switching design from a Fourier-Fisher perspective}
\label{sec:fourierFisherTheory}

Even under the simplifications performed on the ambiguity function, the sequence design involves integrating over the angle and Doppler domains as the base of an objective function. To further reduce computational complexity, the optimization process needs a complete review starting from the theory that describes estimation accuracy and ambiguity.


This section first performs a description of the estimation accuracy of the MPC parameters in Sect.\,\ref{ss:estimationBounds}. Sect.\,\ref{ss:fisherStep} proposes objective functions $J(\boldsymbol{\eta})$ that locally optimize for estimation accuracy. As the optimization for accuracy is shown to be local and can be hampered by ambiguities in the estimation, Sect.\, \ref{ss:fourierStep} proposes an objective function $J_0(\boldsymbol{\eta})$ to globally minimize ambiguity stochastically in the estimation before moving to estimation accuracy optimization. The resulting successive optimization process and the corresponding FF algorithms proposed in Sect.\,\ref{ss:ffSequences} efficiently split the sequence design problem. Sect.\,\ref{ss:complexityFF} analyzes the complexity of the proposed FF method. Sect.\,\ref{ss:ffWidebandExtension} finally outlines how to extend the proposed method to wideband scenarios where the variation in the antenna frequency responses cannot be neglected.

\subsection{Estimation Bounds on Angles and Doppler}
\label{ss:estimationBounds}

Without loss of essence, let us consider a narrowband single-path channel that is measured for one snapshot with a single-polarized TX array and a single-polarized RX array. The received signal $\mathbf{y}\in \mathbb{C}^{M_{\text{TR}}}$ can be written as
\begin{equation} \label{eq:modelSinglePolarization}
\mathbf{y} = \mathbf{s} (\boldsymbol{\theta}_{\text{sp}}) + \mathbf{n} = \gamma (\mathbf{b}_{\text{T}} \otimes \mathbf{b}_{\text{R}}) \odot \mathbf{a}_{\nu} + \mathbf{n},
\end{equation}
where $\mathbf{n}\in \mathbb{C}^{M_{\text{TR}}}$ denotes zero-mean i.i.d circular white Gaussian noise with covariance matrix $\mathbf{R}_{nn}=\sigma^2\mathbf{I}_{M_{\text{TR}}\times M_{\text{TR}}}$, $\gamma = re^{j\psi}$ is the complex amplitude of the path, and $\mathbf{b}_{\text{T}} \in \mathbb{C}^{M_{\text{T}}}$, $\mathbf{b}_{\text{R}} \in \mathbb{C}^{M_{\text{R}}}$ are array response vectors for a single polarization, dependent on the directions of departure $\varphi_{\text{T}}$, $\vartheta_{\text{T}}$, and the directions of arrival $\varphi_{\text{R}}$, $\vartheta_{\text{R}}$, respectively. The vector $\boldsymbol{\theta}_{\text{sp}}$ contains all the propagation path parameters to be estimated. More precisely, $\boldsymbol{\theta}_{\text{sp}} = \begin{bmatrix}\vartheta_{\text{T}} &\varphi_{\text{T}} &\vartheta_{\text{R}} &\varphi_{\text{R}} &\nu &r & \psi\end{bmatrix}^{\text{T}}$. Note that $\mathbf{y} \sim \mathcal{N}(\mathbf{s}(\boldsymbol{\theta}_{\text{sp}}),\mathbf{R}_{nn})$. The variances of any unbiased estimator $\hat{\boldsymbol{\theta}}_{\text{sp}}$ can be then bounded by the Cramér–Rao lower bound (CRLB) \cite{kay1993fundamentals} as $\text{var}([\hat{\boldsymbol{\theta}}_{\text{sp}}]_i) \geq \text{CRLB}([\boldsymbol{\theta}_{\text{sp}}]_i)$, where
\begin{equation} \label{eq:crlbGeneral} 
 \text{CRLB}([\boldsymbol{\theta}_{\text{sp}}]_{i})=[\boldsymbol{F}^{-1}(\boldsymbol{\theta}_{\text{sp}})]_{ii},
\end{equation}
with $\boldsymbol{F}$ being the Fisher information matrix (FIM) of $\boldsymbol{\theta}_{\text{sp}}$. For a multivariate normal distribution with such a covariance matrix structure, the Slepian-Bangs formula \cite{slepian1954estimation,bangs1971array,stoica1997introduction} can be used to express the FIM as
\begin{equation} \label{eq:fim}
\mathbf{F}(\boldsymbol{\theta}_{\text{sp}})=\frac{2}{\sigma^2}\Re \{ \mathbf{D}(\boldsymbol{\theta}_{\text{sp}})^{\text{H}}\cdot \mathbf{D}(\boldsymbol{\theta}_{\text{sp}})  \},
\end{equation}
where
\begin{equation} \label{eq:jacobianChannel}
\boldsymbol{D}(\boldsymbol{\theta}_{\text{sp}}) = \frac{\partial}{\partial \boldsymbol{\theta}_{\text{sp}}} \mathbf{s}(\boldsymbol{\theta}_{\text{sp}})
\end{equation}
is the Jacobian of the transmitted signal. Under this construction, the FIM is a Hermitian positive definite matrix that can be decomposed into $\mathbf{F} = \mathbf{F}_D + \mathbf{F}_H$, where $\mathbf{F}_D = \mathbf{F} \odot \mathbf{I}_{n}$ is a diagonal positive definite matrix and $\mathbf{I}_{n} \in \mathbb{C}^{n \times n}$ is the identity matrix. Sect.\,\ref{ss:fisherStep} finds the optimization function focused on estimator variances, i.e. estimation accuracy. The following mathematical results are needed beforehand.

\begin{lemma}
\label{lemma:csPosDef}
Let $a_k, b_k > 0$ for any $k \in \mathbb{N}^+$ be real numbers and $\sum_k{b_k^2} = 1$. Then,
\begin{equation*}
    \sum_k \frac{b_k^2}{a_k^2} \geq \frac{1}{\sum_k a_k^2 b_k^2}.
\end{equation*}
\end{lemma}
\begin{proof}
See Appendix \ref{app:proofCSPosDef}.
\end{proof}

\begin{theorem}
\label{theorem:fisherCrossElements}
Let $\mathbf{A} = \mathbf{B} + \mathbf{C}$, with $\mathbf{A} \in \mathbb{C}^{n \times n}$ Hermitian positive semidefinite, and $\mathbf{B} = \mathbf{A} \odot \mathbf{I}_{n}$ diagonal, where $\mathbf{I}_{n} \in \mathbb{C}^{n \times n}$ is the identity matrix. Then, $\left[ \mathbf{A}^{-1} \right]_{ii} \geq \left[ \mathbf{B}^{-1} \right]_{ii}$ for any $i = 1,\dots,n$. The equality holds iff the off-diagonal elements of $\mathbf{A}$ are zeros.
\end{theorem}
\begin{proof}
See Appendix \ref{app:proofFisherCrossElements}.
\end{proof}

\subsection{Fisher Step}
\label{ss:fisherStep}

When applying Theorem \ref{theorem:fisherCrossElements} to (\ref{eq:crlbGeneral}), it is clear that forcing the off-diagonal elements of $\mathbf{F}$ to be zero can improve the precision of the estimators of $\boldsymbol{\theta}_{\text{sp}}$ given that the diagonal elements of $\mathbf{F}$ remain stable. In such a case, the precision of the estimators can be bounded by $\displaystyle \text{var}([\hat{\boldsymbol{\theta}}_{\text{sp}}]_i) \geq \frac{1}{[\mathbf{F}]_{ii}}$. More concisely, the optimization problem that addresses estimation precision can be expressed as
\begin{equation}
    \min_{\boldsymbol{\eta} \in \mathfrak{N}_{M_{\text{TR}}}} J\left( \boldsymbol{\eta} \right),
\end{equation}
where $J(\cdot)$ is a cost function that takes the off-diagonal elements of the FIM that are dependent on the switching sequence $\boldsymbol{\eta}$ as input. Since the precision enhancement foundations are the FIM elements, this procedure is referred to as the Fisher step.\\ 

\subsubsection{General case}

With help of the EADF \cite{landmann2004efficient,richter2005estimation}, the array response vector of an array with an arbitrary geometry can be expressed as $\mathbf{b}_{\text{T}/\text{R}} = \mathbf{G}_{\text{T}/\text{R}} \cdot \left( \boldsymbol{\beta}_{\varphi_{\text{T}/\text{R}}} \otimes \boldsymbol{\beta}_{\vartheta_{\text{T}/\text{R}}} \right)$,  where $\mathbf{G}_{\text{T}/\text{R}} \in \mathbb{C}^{M_{\text{T/R}} \times A_{\varphi_{\text{T}/\text{R}}} A_{\vartheta_{\text{T}/\text{R}}}}$ contains the EADFs of the TX/RX antenna elements, $\boldsymbol{\beta}_{\varphi_{\text{T}/\text{R}}} = e^{j \varphi_{\text{T}/\text{R}} \boldsymbol{\alpha}_{\varphi_{\text{T}/\text{R}}}}$, $\boldsymbol{\beta}_{\vartheta_{\text{T}/\text{R}}} = e^{j \vartheta_{\text{T}/\text{R}} \boldsymbol{\alpha}_{\vartheta_{\text{T}/\text{R}}}}$ are phase vectors to recover the array response from the EADFs, with $\boldsymbol{\alpha}_{\varphi_{\text{T}/\text{R}}} \in \mathbb{Q}^{A_{\varphi_{\text{T}/\text{R}}}}, \boldsymbol{\alpha}_{\vartheta_{\text{T}/\text{R}}} \in \mathbb{Q}^{A_{\vartheta_{\text{T}/\text{R}}}}$ being angular frequencies, i.e., $\boldsymbol{\alpha}_{\varphi_{\text{T}/\text{R}}} = \left[ -\frac{A_{\varphi_{\text{T}/\text{R}}}-1}{2}, \dots, \frac{A_{\varphi_{\text{T}/\text{R}}}-1}{2} \right]^{\text{T}}$, $\boldsymbol{\alpha}_{\vartheta_{\text{T}/\text{R}}} = \left[ -\frac{A_{\vartheta_{\text{T}/\text{R}}}-1}{2}, \dots, \frac{A_{\vartheta_{\text{T}/\text{R}}}-1}{2} \right]^{\text{T}}$, and $A_{\varphi_{\text{T}/\text{R}}}$ and $A_{\vartheta_{\text{T}/\text{R}}}$ the numbers of measured azimuth and elevation points.

The diagonal elements of the FIM for different directions of departure/arrival are
\begin{equation}
\begin{aligned}
    [\mathbf{F}]_{\varphi_{\text{T/R}}\varphi_{\text{T/R}}} = \frac{2r^2}{\sigma^2} \Re &\left\{ \left( \left[ \boldsymbol{\beta}_{\varphi_{\text{T/R}}}^{\hermitian} \odot \boldsymbol{\alpha}_{\varphi_{\text{T/R}}}^{\hermitian} \right] \otimes \boldsymbol{\beta}_{\vartheta_{\text{T/R}}}^{\hermitian} \right) \cdot \mathbf{G}_{\text{T/R}}^{\hermitian} \right.\\
    &\left.\ \cdot \mathbf{G}_{\text{T/R}} \cdot \left( \left[ \boldsymbol{\beta}_{\varphi_{\text{T/R}}} \odot \boldsymbol{\alpha}_{\varphi_{\text{T/R}}} \right] \otimes \boldsymbol{\beta}_{\vartheta_{\text{T/R}}} \right) \right\},
\end{aligned}
\end{equation}
\begin{equation}
\begin{aligned}
    [\mathbf{F}]_{\vartheta_{\text{T/R}}\vartheta_{\text{T/R}}} = \frac{2r^2}{\sigma^2} \Re &\left\{ \left( \boldsymbol{\beta}_{\varphi_{\text{T/R}}}^{\hermitian} \otimes \left[ \boldsymbol{\beta}_{\vartheta_{\text{T/R}}}^{\hermitian} \odot \boldsymbol{\alpha}_{\vartheta_{\text{T/R}}}^{\hermitian} \right] \right) \cdot \mathbf{G}_{\text{T/R}}^{\hermitian} \right.\\
    &\left.\ \cdot \mathbf{G}_{\text{T/R}} \cdot \left( \boldsymbol{\beta}_{\varphi_{\text{T/R}}} \otimes \left[ \boldsymbol{\beta}_{\vartheta_{\text{T/R}}} \odot \boldsymbol{\alpha}_{\vartheta_{\text{T/R}}} \right] \right) \right\}.
\end{aligned}
\end{equation}
The FIM diagonal entry for Doppler is
\begin{equation} \label{eq:fimDiagonalDopplerGeneral}
    [\mathbf{F}]_{\nu\nu} = 8 \cdot \left(\frac{r\pi ||\boldsymbol{\eta}||}{\sigma}\right)^2.
\end{equation}
The FIM off-diagonal elements for cross Doppler-angles are
\begin{equation}
\begin{aligned}
    [\mathbf{F}]_{\nu\varphi_{\text{T}}} = \frac{4\pi r^2}{\sigma^2} \Re &\left\{ \left[ \left( \mathbf{b}_{\text{T}} \otimes \mathbf{1}_{\text{M}_{\text{R}}} \right) \odot \boldsymbol{\eta} \right]^{\hermitian} \right.\\
    &\left.\ \cdot \left( \left[ \mathbf{G}_{\text{T}} \cdot \left( \left[ \boldsymbol{\beta}_{\varphi_{\text{T}}} \odot \boldsymbol{\alpha}_{\varphi_{\text{T}}} \right] \otimes \boldsymbol{\beta}_{\vartheta_{\text{T}}} \right) \right] \otimes \mathbf{1}_{\text{M}_{\text{R}}} \right) \right\},
\end{aligned}
\end{equation}
\begin{equation}
\begin{aligned}
    [\mathbf{F}]_{\nu\vartheta_{\text{T}}} = \frac{4\pi r^2}{\sigma^2} \Re &\left\{ \left[ \left( \mathbf{b}_{\text{T}} \otimes \mathbf{1}_{\text{M}_{\text{R}}} \right) \odot \boldsymbol{\eta} \right]^{\hermitian} \right.\\
    &\left.\ \cdot \left( \left[ \mathbf{G}_{\text{T}} \cdot \left( \boldsymbol{\beta}_{\varphi_{\text{T}}} \otimes \left[ \boldsymbol{\beta}_{\vartheta_{\text{T}}} \odot \boldsymbol{\alpha}_{\vartheta_{\text{T}}} \right] \right) \right] \otimes \mathbf{1}_{\text{M}_{\text{R}}} \right) \right\},
\end{aligned}
\end{equation}
\begin{equation}
\begin{aligned}
    [\mathbf{F}]_{\nu\varphi_{\text{R}}} = \frac{4\pi r^2}{\sigma^2} \Re &\left\{ \left[ \left( \mathbf{1}_{\text{M}_{\text{T}}} \otimes \mathbf{b}_{\text{R}} \right) \odot \boldsymbol{\eta} \right]^{\hermitian} \right.\\
    &\left.\ \cdot \left( \mathbf{1}_{\text{M}_{\text{T}}} \otimes \left[ \mathbf{G}_{\text{R}} \cdot \left( \left[ \boldsymbol{\beta}_{\varphi_{\text{R}}} \odot \boldsymbol{\alpha}_{\varphi_{\text{R}}} \right] \otimes \boldsymbol{\beta}_{\vartheta_{\text{R}}} \right) \right] \right) \right\},
\end{aligned}
\end{equation}
\begin{equation}
\begin{aligned}
    [\mathbf{F}]_{\nu\vartheta_{\text{R}}} = \frac{4\pi r^2}{\sigma^2} \Re &\left\{ \left[ \left( \mathbf{1}_{\text{M}_{\text{T}}} \otimes \mathbf{b}_{\text{R}} \right) \odot \boldsymbol{\eta} \right]^{\hermitian} \right.\\
    &\left.\ \cdot \left( \mathbf{1}_{\text{M}_{\text{T}}} \otimes \left[ \mathbf{G}_{\text{R}} \cdot \left( \boldsymbol{\beta}_{\varphi_{\text{R}}} \otimes \left[ \boldsymbol{\beta}_{\vartheta_{\text{R}}} \odot \boldsymbol{\alpha}_{\vartheta_{\text{R}}} \right] \right) \right] \right) \right\}.
\end{aligned}
\end{equation}
For a more detailed derivation, please refer to Appendix \ref{app:fimEntriesGeneral}. Note from Appendix \ref{app:fimEntriesGeneral} that the parameter $\boldsymbol{\eta}$ is present in the diagonal entries of the FIM as subject to a norm, and is present in all of the off-diagonal entries containing Doppler. Recalling (\ref{eq:switchingVectorCanonical}), the structure of $\boldsymbol{\eta}$ is composed of a fixed $\mathbf{m}_{\text{TR}}$ and $\Delta t$, and a configurable permutation matrix $\mathbf{P}_\pi$. This implies that $||\boldsymbol{\eta}||$ does not change despite changes in $\mathbf{P}_\pi$, i.e., changes in $\boldsymbol{\eta}$. Hence, there is no dependence in the diagonal elements of the FIM with respect to $\boldsymbol{\eta}$. No dependence on $\boldsymbol{\eta}$ was found in the rest of the off-diagonal entries of the FIM. Therefore, the CRLB can be minimized by solving
\begin{equation}
\begin{aligned}
\label{eq:minCRLBGeneral}
    \min_{\boldsymbol{\eta}} J(\boldsymbol{\eta}) = \min_{\boldsymbol{\eta}} \quad &f_1([\mathbf{F}]_{\nu\varphi_{\text{T}}}) + f_2([\mathbf{F}]_{\nu\vartheta_{\text{T}}})\\
    &+ f_3([\mathbf{F}]_{\nu\varphi_{\text{R}}}) + f_4([\mathbf{F}]_{\nu\vartheta_{\text{R}}}).
\end{aligned}
\end{equation}
where $f_1(\cdot)$, $f_2(\cdot)$, $f_3(\cdot)$, $f_4(\cdot)$ are functions that quantify a partial cost from each of the off-diagonal elements of the FIM dependent on $\boldsymbol{\eta}$. The proposed functions are
\begin{equation}
\begin{aligned}
\label{eq:fisherPartialCosts}
    f_1([\mathbf{F}]_{\nu\varphi_{\text{T}}}) &= \max_{\vartheta_{\text{T}},\varphi_{\text{T}}} \left| [\mathbf{F}]_{\nu\varphi_{\text{T}}} \right| \\
    f_2([\mathbf{F}]_{\nu\vartheta_{\text{T}}}) &= \max_{\vartheta_{\text{T}},\varphi_{\text{T}}} \left| [\mathbf{F}]_{\nu\vartheta_{\text{T}}} \right| \\
    f_3([\mathbf{F}]_{\nu\varphi_{\text{R}}}) &= \max_{\vartheta_{\text{R}},\varphi_{\text{R}}} \left| [\mathbf{F}]_{\nu\varphi_{\text{R}}} \right| \\
    f_4([\mathbf{F}]_{\nu\vartheta_{\text{R}}}) &= \max_{\vartheta_{\text{R}},\varphi_{\text{R}}} \left| [\mathbf{F}]_{\nu\vartheta_{\text{R}}} \right|.
\end{aligned}
\end{equation}
The absolute values of the off-diagonal elements are taken to quantify their magnitudes. The off-diagonal elements show a dependence on the vectors $\boldsymbol{\beta}_{\vartheta_{\text{T}}}$, $\boldsymbol{\beta}_{\varphi_{\text{T}}}$, $\boldsymbol{\beta}_{\vartheta_{\text{R}}}$, $\boldsymbol{\beta}_{\varphi_{\text{R}}}$, i.e., on the angles $\vartheta_{\text{T}}$, $\varphi_{\text{T}}$, $\vartheta_{\text{R}}$, $\varphi_{\text{R}}$. As observed in the derivation from (\ref{eq:fimDopplerAOD}), no dependence on Doppler $\nu$ was found for these elements. Therefore, the functions take the highest magnitude over angles that an off-diagonal element can have as its output partial cost.

\subsubsection{Locality of Fisher information}

The FIM can be interpreted as the Hessian of the minimized relative entropy between the true data distribution, with parameters $\boldsymbol{\theta}$, and the data distribution with approximated parameters $\boldsymbol{\theta}'$ \cite{kay1993fundamentals,gourieroux1995statistics}. This is because the relative entropy is minimized for $\boldsymbol{\theta}' = \boldsymbol{\theta}$, and its curvature in the vicinity of a parameter value $\boldsymbol{\theta}$ is given by minus the expectation of the curvature of the log-likelihood function evaluated at that parameter value. This shows that the Fisher information metric is local and is limited to values of $\boldsymbol{\theta}'$ that are close to the true ones. When performing parametric estimation, the Fisher information thus describes the precision of the main estimation lobe only, not the side lobes present in the rest of the search space. That is, the ambiguity in the estimation when using switching sequences is not fully handled by the Fisher step. This suggests that a prior step should minimize ambiguity and provide an initial sequence, which can be refined later to improve the precision of its associated main estimation peak under MLE, by solving (\ref{eq:minCRLBGeneral}). The prior step is now described  in Sect.\,\ref{ss:fourierStep}, referred to as the Fourier step, since it works with the Fourier spectrum of switching sequences.

\subsection{Fourier Step}
\label{ss:fourierStep}

The ambiguity function described in \cite{schmidt1986multiple} and the respective development for switching sequences in \cite{pedersen2008optimization,wang2019channel} can consider \textit{any} $\boldsymbol{\theta}'$ in the search space limited by the angular range of the antenna arrays and a Doppler range bounded by half the antenna switching rate. The authors also clarified that the side lobes of the ambiguity can be minimized by optimizing the antenna switching sequence. (\ref{eq:phaseVectorDoppler}) shows that the choice of antenna switching sequence influences one of the phase vectors that make up the basis matrix of the wireless channel. Moreover, \cite{avital2021design} showed that there is a clear link between the MLE and the Fourier transform in the parametric estimation problem and that the design of switching sequences can leverage this link. This happens because wireless systems are working with sinusoidal signals embedded in noise. The received signal can thus be modeled as a sum of sinusoidal components whose periodic behavior allows for a clearer representation in the frequency domain. In other words, the repetition patterns in the chosen switching sequence appear as the repetition patterns in the basis matrix vectors. In turn, these repetition patterns reflect ambiguities in the estimation and can be well identified in the frequency domain by using the Fourier transform. To achieve an ambiguity-optimal switching sequence, it is essential to find a sequence that minimizes repetition patterns in its content, observed in the dual spectrum as discrete frequencies. Thus, the optimization problem that addresses estimation ambiguity can be expressed as
\begin{equation}
    \min_{\boldsymbol{\eta} \in \mathfrak{N}_{M_{\text{TR}}}} J_0\left( \boldsymbol{\eta} \right),
\end{equation}
where $J_0(\cdot)$ is a cost function that takes a switching sequence $\boldsymbol{\eta}$ as input. A noise-like spectrum with little to no big outlier peaks would be preferable. Therefore, the selection criteria for an initial sequence can be expressed as
\begin{equation} \label{eq:initialCostFunction}
\begin{aligned}
    J_0(\boldsymbol{\eta}) = \text{med}\left[ \mathscr{F} \left\{ \frac{\boldsymbol{\eta}}{\Delta t} \right\} \right],
\end{aligned}
\end{equation}
where $\text{med}[\mathbf{a}]$ is the empirical median of the samples in the vector $\mathbf{a}$. A brute-force approach to this problem would yield the expression
\begin{equation} \label{eq:bruteForceFourierStep}
\begin{aligned}
    \boldsymbol{\eta}_0 = \arg \min_{\boldsymbol{\eta} \in \mathfrak{N}_{M_{\text{TR}}}} \text{med}\left[ \mathscr{F} \left\{ \frac{\boldsymbol{\eta}}{\Delta t} \right\} \right].
\end{aligned}
\end{equation}

The number of sequences to score and hence the time complexity of (\ref{eq:bruteForceFourierStep}) increases with the factorial of the number of antenna pair combinations $M_{\text{TR}}$. Since the population becomes prohibitively large, a reliable way of reducing the time complexity in a controlled manner is to use Cochran's sampling size formula \cite{cochran1977sampling}. The formula estimates a required sample size -- out of the entire population of sequences $M_{\text{TR}}!$ -- while allowing precise probabilistic control on performance degradation with respect to the global optimum. Since every switching sequence has an associated cost given by (\ref{eq:initialCostFunction}), it is possible to sort a population sample in percentiles and infer percentile ranks (percentage of population whose score is less than the score of an individual sample) for the entire population. Associating the percentile rank with the Fourier cost in (\ref{eq:initialCostFunction}), the lowest-ranked sequence is then selected as an initial sequence. The inference is done with a configurable margin of error and confidence level. The margin of error $d$ represents the maximum deviation of any estimated percentile rank from the true rank. The confidence level $CL$ represents the probability that the estimated percentile rank is not deviating from the true rank by more than the margin of error (i.e. the confidence we have in the estimated rank interval of including the true rank).  Cochran's sampling size formula estimates the required sample size $n_0$ of a large population of switching sequences to be
\begin{equation} \label{eq:cochransSampleSize}
n_0 = \frac{t^2p(1-p)}{d^2},
\end{equation}
where $t = F^{-1}\left\{ 1-\frac{1-CL}{2} \right\}$ gives the z-value corresponding to the desired confidence level $CL$ assuming the percentile ranks are normally distributed, $F^{-1}\{\cdot\}$ is the inverse cumulative distribution function, and $p=0.5$ accounts for no prior information on the percentile rank of any sequence (50\% chance of being in any rank). For example, to find a switching sequence whose Fourier cost sits in the bottom 5\% of all sequences (i.e. the 5\% best switching sequences in terms of ambiguity), with 99\% of success, the margin of error of the percentile rank is set to $d = 0.05$, the confidence level is set to $CL = 0.99$, the sample size $n_0$ is calculated according to (\ref{eq:cochransSampleSize}), and the switching sequence with the lowest Fourier cost (i.e. percentile rank 0\%) is selected from the population sample of sequences. The Fourier step of the FF design can be compressed as
\begin{equation} \label{eq:initialSequence}
\begin{aligned}
    \boldsymbol{\eta}_0 = \arg \min_{\boldsymbol{\eta} \in \mathfrak{M}_{M_{\text{TR}}}} \quad J_0(\boldsymbol{\eta}),
\end{aligned}
\end{equation}
where $\mathfrak{M}_{M_{\text{TR}}} = \left\{ \boldsymbol{\eta}: \boldsymbol{\eta} \in_{\text{R}} \mathfrak{N}_{M_{\text{TR}}} \right\}$, $\left| \mathfrak{M}_{M_{\text{TR}}} \right| = n_0$ is the cardinality of the set $\mathfrak{M}_{M_{\text{TR}}}$, and the notation ``$\in_{\text{R}}$'' expresses a random pick of an element from a set. From here, any iterative algorithm can find the solution to (\ref{eq:minCRLBGeneral}).

\subsection{Fourier-Fisher Switching Sequences}
\label{ss:ffSequences}

In this subsection, we show a two-step optimization scheme for switching sequences. First, the Fourier step explained in Sect.\,\ref{ss:fourierStep} minimizes the side lobes in the channel parametric estimation by eliminating repetition patterns in the sequences under test. This is achieved by evaluating the Fourier spectrum of a representative sample size of switching sequences. Then, the Fisher step explained in Sect.\,\ref{ss:fisherStep} minimizes the width of the main estimation lobe, related to the accuracy with respect to the set of ground truth parameters. This is achieved by minimizing a cost function dependent on the off-diagonal elements of the FIM. We refer to the sequences found with this method as Fourier-Fisher sequences.

\begin{algorithm}
\caption{The simulated-annealing algorithm to solve the optimization problem (\ref{eq:minCRLBGeneral}). Adapted from \cite{al2023hybrid}.}
\label{algorithm:SA}
\begin{tabular}{l}
1: Initialize $\boldsymbol{\eta}$ according to (\ref{eq:initialSequence}), the temperature $T=T_0$,\\\quad and $\alpha = \alpha_0$ \\
2: \textbf{while} $k \leq k_{\text{max}}$ \textbf{do}\\
3: \quad $\boldsymbol{\eta}' = \text{Update}(\boldsymbol{\eta})$\\
4: \quad $\textbf{if } \exp([(J(\mathbf{F}(\boldsymbol{\eta}))-J(\mathbf{F}(\boldsymbol{\eta}')))/T])>\text{random}(0,1) \textbf{ then}$\\
5: \qquad $\boldsymbol{\eta} = \boldsymbol{\eta'}$\\ 
6: \quad \textbf{end if}\\
7: \quad $T=\alpha T$\\
8: \textbf{end while}\\
9: \textbf{return} $\boldsymbol{\eta}$\\
\end{tabular}
\end{algorithm}

An algorithm is presented in Algorithm \ref{algorithm:SA} for the design of FF sequences. The algorithm starts by initializing the switching sequence given the Fourier step expressed in (\ref{eq:initialSequence}). Then the Fisher-based metric is optimized with simulated annealing. This caters to a very large number of antenna pairs in the system. The parameters determining the sampling size for the Fourier step or the simulated annealing related parameters for Algorithm \ref{algorithm:SA} are implementation specific and depend on the performance requirements of the switching sequence. Another algorithmic approach could use the divide-and-conquer paradigm as an example. That is, the refinement process is not limited to this algorithm, and readers are encouraged to come up with their own solutions.

\subsection{Computational complexity of antenna switching design based on Fourier-Fisher theory}\label{ss:complexityFF}

The entire Fourier step in (\ref{eq:initialSequence}) attains a complexity

\begin{equation}
\begin{aligned}
C_{\text{Fo}} &= \mathcal{O}\left( n_0 \left( M_{\text{TR}} + M_{\text{TR}} \log{M_{\text{TR}}} + M_{\text{TR}} \right) \right)\\
&= \mathcal{O}\left( n_0 M_{\text{TR}} \log{M_{\text{TR}}} \right).
\end{aligned}
\end{equation}

The general cost function in (\ref{eq:minCRLBGeneral}), used in the Fisher step, attains a complexity
\begin{equation}
\begin{aligned}
C_{\text{Fi}} &= \mathcal{O}\left( M_{\vartheta_{\text{T}}} M_{\varphi_{\text{T}}} \left( 5 M_{\text{TR}} + M_{\text{T}} A_{\varphi_{\text{T}}} A_{\vartheta_{\text{T}}} + A_{\varphi_{\text{T}}} A_{\vartheta_{\text{T}}} + A_{\varphi_{\text{T}}} \right) \right)\\
&\, + \mathcal{O}\left( M_{\vartheta_{\text{T}}} M_{\varphi_{\text{T}}} \left( 5 M_{\text{TR}} + M_{\text{T}} A_{\varphi_{\text{T}}} A_{\vartheta_{\text{T}}} + A_{\varphi_{\text{T}}} A_{\vartheta_{\text{T}}} + A_{\vartheta_{\text{T}}} \right) \right)\\
&\, + \mathcal{O}\left( M_{\vartheta_{\text{R}}} M_{\varphi_{\text{R}}} \left( 5 M_{\text{TR}} + M_{\text{R}} A_{\varphi_{\text{R}}} A_{\vartheta_{\text{R}}} + A_{\varphi_{\text{R}}} A_{\vartheta_{\text{R}}} + A_{\varphi_{\text{R}}} \right) \right)\\
&\, + \mathcal{O}\left( M_{\vartheta_{\text{R}}} M_{\varphi_{\text{R}}} \left( 5 M_{\text{TR}} + M_{\text{R}} A_{\varphi_{\text{R}}} A_{\vartheta_{\text{R}}} + A_{\varphi_{\text{R}}} A_{\vartheta_{\text{R}}} + A_{\vartheta_{\text{R}}} \right) \right)\\
&\, + \mathcal{O}(1)\\
&= \mathcal{O}\left( M_{\vartheta_{\text{T}}} M_{\varphi_{\text{T}}} \left( M_{\text{TR}} + M_{\text{T}} A_{\varphi_{\text{T}}} A_{\vartheta_{\text{T}}} \right) \right)\\
&\, + \mathcal{O}\left( M_{\vartheta_{\text{R}}} M_{\varphi_{\text{R}}} \left( M_{\text{TR}} + M_{\text{R}} A_{\varphi_{\text{R}}} A_{\vartheta_{\text{R}}} \right) \right).
\end{aligned}
\end{equation}

The implementation of the Fisher step through Algorithm \ref{algorithm:SA} is a metaheuristic whose iterations attain a complexity dictated by the choice of cost function. Hence, its time complexity will be analyzed via simulations in Sect.\,\ref{sec:simulationResults}.

\subsection{Extension to wideband scenarios}
\label{ss:ffWidebandExtension}

The antenna switching design based on FF sequences can be easily adapted to wideband scenarios by dropping the assumption of flat antenna array response. This can be realized, for instance, by varying the polarimetric antenna responses across $M_f$ points. The Fisher cost function (\ref{eq:minCRLBGeneral}) is then calculated for each frequency point, and then added to a total cost function for all frequency points. The complexity of the cost function grows linearly with $M_f$ in this case. A more efficient method is to calculate a composite radiation pattern of each antenna. This can be done by performing an inverse Fourier transform on the antenna's frequency response at each angle and using the complex amplitude of the highest peak in the transformed domain, i.e. the ``delay'' domain, as the composite pattern. This could be seen as averaging the effect of all frequency points. The complexity of the cost function does not grow.

\section{Simulations with Realistic Array Responses}
\label{sec:simulationResults}


\begin{table}
\begin{center}
	\caption{Simulation setup.}
	\label{tab:simulationSetup}
	\begin{tabular}{|c|c|}
		
		\hline	
        \textbf{Processor} & Intel(R) Core(TM) i7-8565U CPU\\\hline
        \textbf{RAM} & $16$\,GB\\\hline
        \textbf{Operative system} & Windows 10 x64\\\hline
		\textbf{Carrier frequency}	& $28$\,GHz\\\hline
		\textbf{TX radio head} & ULA with 64 elements\\\hline
        \textbf{TX element type} & Patch antenna, vertically polarized,\\
        & realistic antenna pattern according\\
        & to the measurement in \cite{cai2024switched}\\\hline
		\textbf{TX switching rate} & $18.8$\,$\mu$s\\\hline
        \textbf{RX radio head} & Single antenna\\\hline
        \textbf{RX element type} & Isotropic, vertically polarized\\\hline
        \textbf{Azimuth true value} & $90$\degrees\\\hline
        \textbf{Azimuth search space} & $[-180,180]$\degrees\\\hline
        \textbf{SNR range} & $[-20,10]$\,dB\\\hline
        \textbf{SNR points} & $201$\\\hline
        \textbf{Simulations per SNR point} & $10000$\\\hline
		\textbf{Total simulations per sequence} & $2010000$\\\hline
        \textbf{Sequence 1} & Trivial\\\hline
        \textbf{Sequence 2} & Ambiguity \cite{wang2019channel}, $\mathcal{P} = 6$\\\hline
        \textbf{Sequence 3} & FF with realistic TX pattern, \\&confidence level 99\%, \\&error margin 1\%\\\hline
        \textbf{Sequence 4} & FF with isotropic TX pattern, \\&confidence level 99\%, \\&error margin 1\%\\\hline
        \textbf{Computation time sequence 1} & None\\\hline
        \textbf{Computation time sequence 2} & $23.122$\,s\\\hline
        \textbf{Computation time sequence 3} & $0.836$\,s total, $0.713$\,s Fisher step\\\hline
        \textbf{Computation time sequence 4} & $0.115$\,s total, $0.002$\,s Fisher step\\\hline
	
	\end{tabular}
\end{center}
\end{table}

The performance of the FF sequences proposed in Sect.\,\ref{ss:ffSequences} was evaluated using Monte Carlo simulations where MLE is used to estimate path parameters for a single path. The simulation setup and mean computation times\footnote{The mean computation times were calculated from a set of 5 independent runs for each sequence.} of the switching sequences under test were collected in Table \ref{tab:simulationSetup}. The simulation consists of a channel sounder with a realistic uniform linear array (ULA) at the TX side and a single isotropic antenna element at the RX side. To retain the essence of the comparison and enhance clarity in the analysis, two simplifications were made. First, all antennas are assumed to be perfectly vertically polarized. Second, only azimuth of departure (simply referred to as azimuth hereafter) and Doppler were left unknown in the estimation algorithm, and the path elevation was assumed to be 0\degrees, i.e., only 1D azimuth estimation was considered. The measured radiation patterns of realistic patch antennas in \cite{cai2024switched} were applied to the elements of the ULA in the simulation. Four different TX switching sequences were tested, namely a trivial sequence, an ambiguity sequence based on \cite{wang2019channel} using the measured radiation pattern of the antenna array, an FF sequence using the measured radiation pattern of the antenna array and an FF sequence assuming a ULA of isotropic antenna elements in place of the realistic ULA, (i.e., not matching but approximating the realistic radiation pattern of the TX array). The selected comparison metric was the RMSE and the logarithm of the MSE. Realistic CRLBs in azimuth and Doppler were also computed as a reference on the best achievable unbiased estimation performance. More than two million simulation runs were performed for each switching sequence across 201 SNR levels, using the same noise realization for a single run across all sequences for a fair comparison.

Since the focus of the estimation was on azimuth and Doppler, the computation complexity of the ambiguity-based and FF switching sequences decreased in line with the simplification of the corresponding cost functions. The complexity of the ambiguity objective function from (\ref{eq:objectiveAmbiguity}) is simplified to $\mathcal{O}\left( M_{\varphi_{\text{T}}} M_{\nu} \mathcal{P} M_{\text{T}} \right)$. The cost function of the realistic optimization problem in (\ref{eq:minCRLBGeneral}) is simplified to
\begin{equation} \label{eq:costGeneralSimplifiedSimulation}
\begin{aligned}
    J(\boldsymbol{\eta}) &= f_1\left( [\mathbf{F}]_{\nu\varphi_{\text{T}}} \right)\\
    &= \max_{\varphi_{\text{T}}} \quad \left| \Re \left\{ \left[ \mathbf{b}_{\text{T}} \odot \boldsymbol{\eta} \right]^{\hermitian} \cdot \mathbf{G}_{\text{T}} \cdot \left[ \boldsymbol{\beta}_{\varphi_{\text{T}}} \odot \boldsymbol{\alpha}_{\varphi_{\text{T}}} \right] \right\} \right|,
\end{aligned}
\end{equation}
where $\mathbf{b}_{\text{T}} = \mathbf{G}_{\text{T}} \cdot \boldsymbol{\beta}_{\varphi_{\text{T}}}$ and $\boldsymbol{\beta}_{\varphi_{\text{T}}}$ is dependent on $\varphi_{\text{T}}$. The complexity of (\ref{eq:costGeneralSimplifiedSimulation}) is then simplified to $\mathcal{O}\left( M_{\varphi_{\text{T}}} M_{\text{T}} A_{\varphi_{\text{T}}} \right)$. Under the assumption of a ULA of isotropic antenna elements, the TX array response vector can be simplified to $\mathbf{b}_{\text{T}} = e^{-j \mathbf{m}_{\text{T}} \mu_{\varphi_{\text{T}}} }$, with $\mu_{\varphi_{\text{T}}} = 2 \pi \frac{d_{\text{r}}}{\lambda} \cos{\varphi_{\text{T}}}$. The off-diagonal FIM element for cross Doppler-azimuth is then $ \displaystyle[\mathbf{F}]_{\nu\varphi_{\text{T}}} = \frac{-4\pi r^2}{\sigma^2} \pdv{\mu_{\varphi_{\text{T}}}}{\varphi_{\text{T}}} \boldsymbol{\eta}^{\hermitian} \mathbf{m}_{\text{T}}$, which can be easily obtained from (\ref{eq:fim}). Therefore, the cost function for the isotropic approximation is further simplified to
\begin{equation}
\begin{aligned}
    J(\boldsymbol{\eta}) = f_1\left( [\mathbf{F}]_{\nu\varphi_{\text{T}}} \right) = \boldsymbol{\eta}^{\hermitian} \cdot \mathbf{m}_{\text{T}},
\end{aligned}
\end{equation}
with a simplified complexity of $\mathcal{O}\left( M_{\text{T}} \right)$.

The corresponding run times to find optimized sequences according to the different schemes under test, as shown in Table \ref{tab:simulationSetup}, reflect the expected results from the theoretical complexities. It is clear that constructing a trivial sequence requires a negligible amount of time, whereas the optimized schemes (both ambiguity and FF) require several steps to output a switching sequence. Moreover, it is remarkable that designing an ambiguity-based switching sequence with measured antenna radiation patterns takes almost 30 times longer than designing an FF sequence with measured antenna radiation patterns and above 200 times longer than designing an FF sequence using isotropic antenna patterns. The difference in computation times shows that FF sequences outperform ambiguity-based sequences by orders of magnitude in terms of computation time. This becomes more evident and pivotal once sequences have to be designed considering both elevation and azimuth estimation and/or with a larger number of antenna elements at the sounder's antenna arrays. It is also worth noting that the Fisher step for the FF sequence design takes around 350 times shorter to run when using theoretical antenna constructions with respect to using measured radiation patterns. This represents a huge increase in computational speeds in cases when the Fisher step is the dominant complexity component in the FF design approach.



\begin{figure}
     \centering
     \begin{subfigure}[b]{0.48\textwidth}
         \centering
         \includegraphics[width=\textwidth]{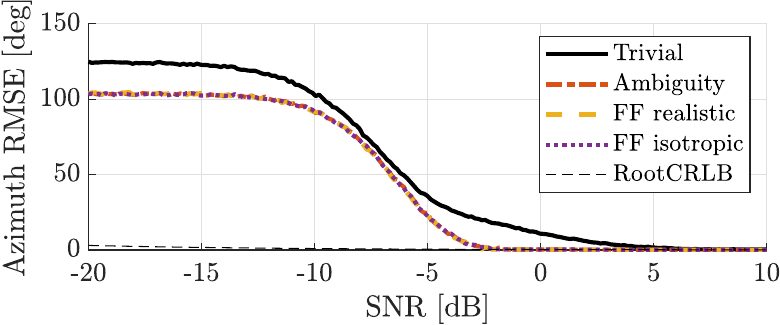}
         \caption{RMSE}
         \label{fig:rmseAzimuth}
     \end{subfigure}
     \hfill
     \begin{subfigure}[b]{0.48\textwidth}
         \centering
         \includegraphics[width=\textwidth]{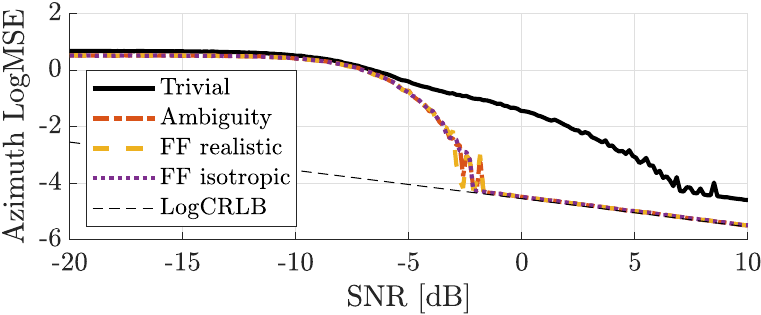}
         \caption{LogMSE}
         \label{fig:logmseAzimuth}
     \end{subfigure}
     \caption{Azimuth estimation error under different sequences. Note that Ambiguity, FF realistic and FF isotropic are basically on top of each other.} \label{fig:errorAOA}
\end{figure}

Fig. \ref{fig:rmseAzimuth} shows the azimuth RMSE for estimation when using the different switching sequences, and the square root of the azimuth CRLB. It is clear that estimation with trivial switching cannot come close in performance to switching sequences optimized for channel sounding, as discussed in \cite{wang2019channel}. However, the trivial sequence can still estimate the true azimuth value following the CRLB with some offset for high SNR levels. This can be explained by the fact that measured radiation patterns are used in the simulations. This variability in the radiation patterns reduces the strength of the ambiguous peaks, rendering them weaker than the main peak corresponding to the true azimuth value. Notice that even though the ambiguous peaks are mitigated, their strength is still comparable to the main true peak. This consequently causes that trivial switching performs worse than all the optimized switching sequences. Furthermore, the estimation performance of the FF sequences under evaluation is comparable to that of the reference ambiguity sequence. When comparing the logarithm of the MSE with the logarithm of the CRLB, as shown in Fig. \ref{fig:logmseAzimuth}, it is clear that all optimized sequences almost achieve the estimation CRLB, with some more pronounced variations occurring around -2\,dB. The variations actually occur across the whole SNR axis but are less noticeable in the rest of the SNR levels, and can be explained by the limitation in the amount of Monte Carlo simulation runs performed. The larger the number of simulation runs performed, the smoother the curves should become. It is interesting to note that the isotropic FF sequence performs equally well as the realistic FF sequence when using realistic radiation patterns for parametric estimation. The geometry of the antenna array helps to achieve such a performance in this case, and this suggests that its geometry can also alleviate the computational complexity of switching sequence optimization for channel sounding.

\begin{figure}
\begin{center}
	\includegraphics[width=0.95\columnwidth]{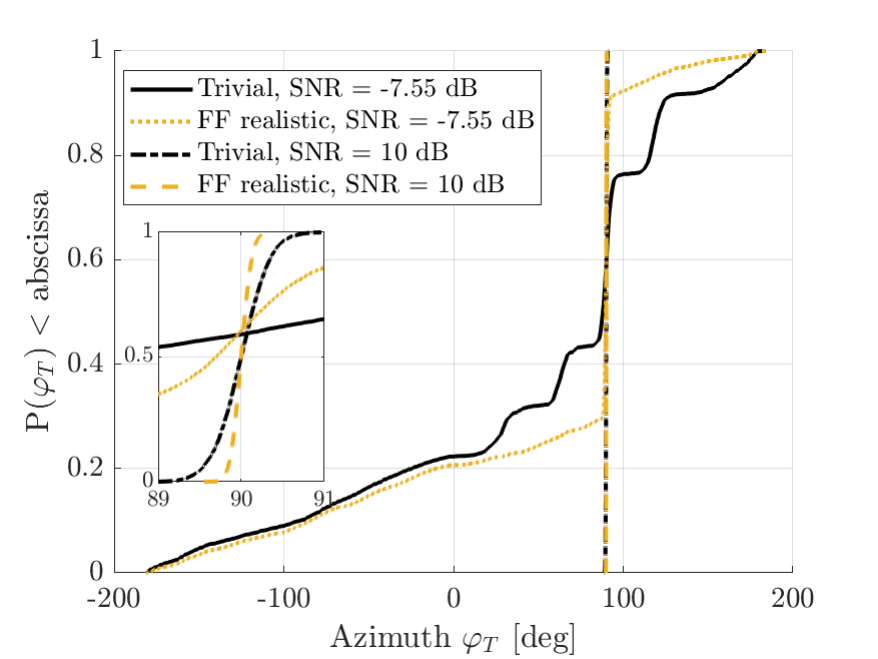}
	\caption{Empirical CDF plot of azimuth estimates for different SNR values.}
	\label{fig:cdfAzimuthComparison}
    \vspace{-4mm}
\end{center}
\end{figure}

\begin{figure}
     \centering
     \begin{subfigure}[b]{0.48\textwidth}
         \centering
         \includegraphics[width=\textwidth]{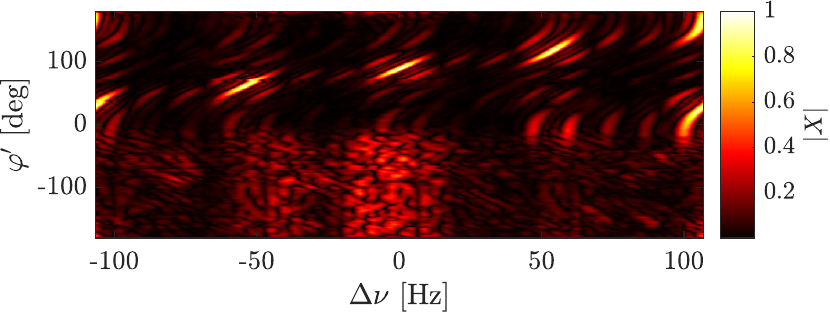}
         \caption{Trivial}
         \label{fig:ambPlotTrivial}
     \end{subfigure}
     \hfill
     \begin{subfigure}[b]{0.48\textwidth}
         \centering
         \includegraphics[width=\textwidth]{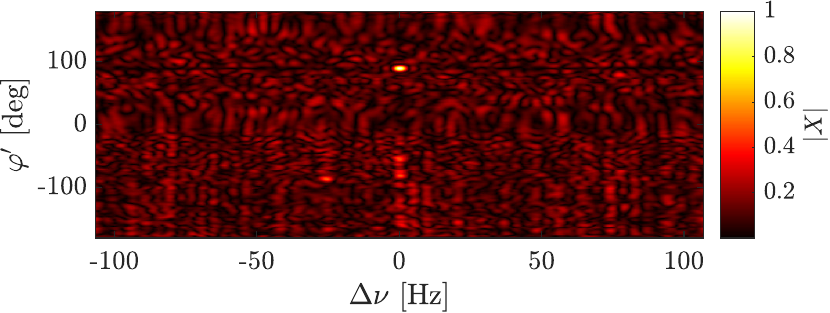}
         \caption{Ambiguity}
         \label{fig:ambPlotAmbiguity}
     \end{subfigure}
     \hfill
     \begin{subfigure}[b]{0.48\textwidth}
         \centering
         \includegraphics[width=\textwidth]{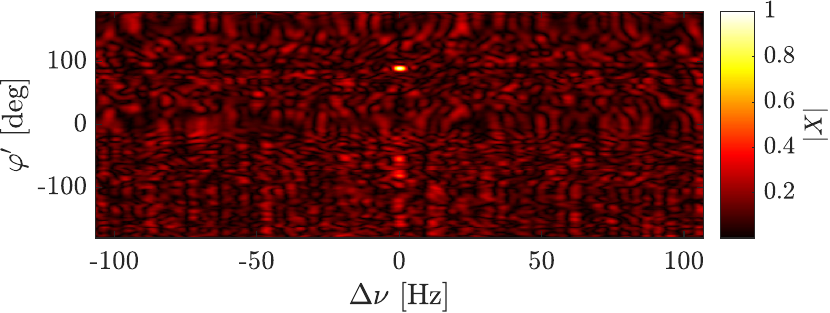}
         \caption{FF realistic}
         \label{fig:ambPlotFFRealistic}
     \end{subfigure}
     \hfill
     \begin{subfigure}[b]{0.48\textwidth}
         \centering
         \includegraphics[width=\textwidth]{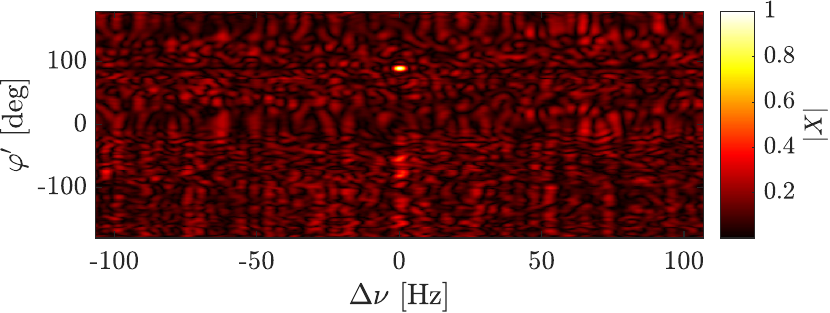}
         \caption{FF isotropic}
         \label{fig:ambPlotFFIsotropic}
     \end{subfigure}
     \caption{Spatio-temporal ambiguity functions for different switching sequences. Note that the radiation patterns of the realistic patch antennas are not isotropic and have a main lobe between $[0,180]\vardegrees$. These features influence the observed ambiguity function behavior.} \label{fig:ambFunctions}
\end{figure}

Having a closer look at the estimation error behavior in Fig. \ref{fig:rmseAzimuth}, it is clearly observed that the performance of trivial switching is comparable to that of all the other switching sequences, when the SNR is low, e.g., below $-8$\,dB. To understand the rationale behind this observed behavior, it is crucial to have a deeper look into the azimuth estimates resulting from the simulations at different SNR levels. Given the significant number of Monte Carlo simulations, an empirical CDF of the azimuth estimates can be computed for every SNR level. Fig. \ref{fig:cdfAzimuthComparison} shows the CDFs of the azimuth estimator under trivial and realistic FF switching for two different SNR values, namely -7.55\,dB and 10\,dB. The different SNR values help show the estimation behavior under moderate and negligible noise levels, respectively. It is clear that the realistic FF sequence exhibits better performance than the trivial sequence for low noise levels, which is characterized by a higher estimate density around the true azimuth value. However, the performance of these two sequences is comparable at moderate noise levels, and the behavior of their CDFs varies significantly. This interesting remark shows the effect of ambiguities in the estimation statistics. As \cite{wang2019channel} mentions, the current design of switching sequences that actively reduce ambiguity effects is spreading the energy of ambiguity side lobes elsewhere in the parameter domains. This is also shown in Fig. \ref{fig:ambFunctions}, where the spatio-temporal ambiguity functions of each sequence are calculated. Therefore, the CDF of the estimates under trivial switching exhibits concentrations of energy around certain azimuth intervals, which basically represent ambiguities. This effect is significantly reduced for the realistic FF sequence, where the energy of the side lobes is spread relatively uniformly across the whole azimuth range outside the main estimation peak, with less concentrated energy levels. However, the higher probability of the estimator falling into neighboring ambiguous peaks for the trivial switching sequence causes the estimated value to be comparable to that of the realistic FF sequence, whose estimator could fall anywhere in the estimation range when the SNR is low. More precisely for Fig. \ref{fig:cdfAzimuthComparison}, the dark solid line shows four ambiguity side bumps at azimuth values of around \{25,60,120,165\}\degrees\ surrounding the main bump at 90\degrees. If another switching sequence increased either the number of bumps or the bump ``height'', so that the dark solid line went below the light dotted line for $\varphi_{\text{T}} < 90\vardegrees$ and above it for $\varphi_{\text{T}} > 90\vardegrees$, the associated RMSE from Fig. \ref{fig:rmseAzimuth} could even go lower than that of FF and ambiguity-based sequences. This suggests that there is potential for designing dynamic switching sequences based on the SNR level that the system experiences. If the estimation performance at lower SNR levels could be increased by spreading the side lobe energy closest to the true parameter values, the resulting sequences would outperform switching sequences that spread this energy evenly across all domain ranges.



\begin{figure}
     \centering
     \begin{subfigure}[b]{0.48\textwidth}
         \centering
         \includegraphics[width=\textwidth]{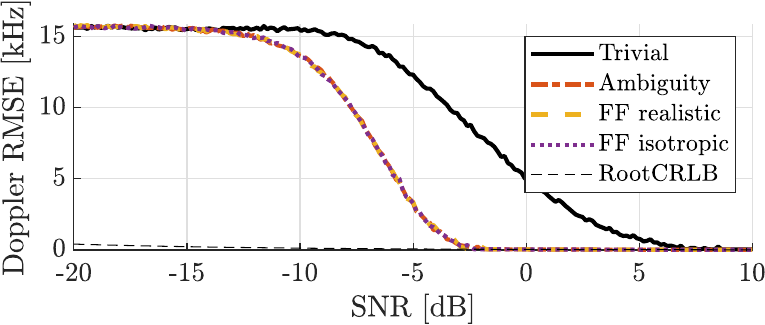}
         \caption{RMSE}
         \label{fig:rmseDoppler}
     \end{subfigure}
     \hfill
     \begin{subfigure}[b]{0.48\textwidth}
         \centering
         \includegraphics[width=\textwidth]{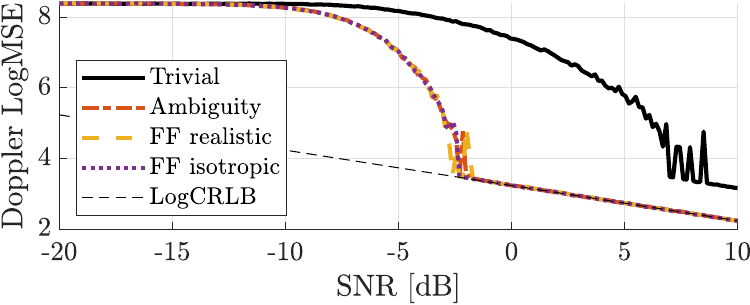}
         \caption{LogMSE}
         \label{fig:logmseDoppler}
     \end{subfigure}
     \caption{Doppler estimation error under different sequences.} \label{fig:errorDoppler}
     \vspace{-3.8mm}
\end{figure}

Fig. \ref{fig:rmseDoppler} shows the Doppler RMSE for estimation under the different sequences. The results in the Doppler domain correlate positively with the results in the azimuth domain. That is, the performance for trivial switching is worse than all the optimized sequences, and both FF switching sequences perform equally well as the baseline ambiguity-based sequence. In addition, all optimized sequences achieve the CRLB at high SNR levels, as shown in Fig. \ref{fig:logmseDoppler}.

\section{Measurement Verification}
\label{sec:measurementResults}

\begin{figure}
     \centering
     \begin{subfigure}{0.54\columnwidth}
         \centering
         \includegraphics[width=\textwidth]{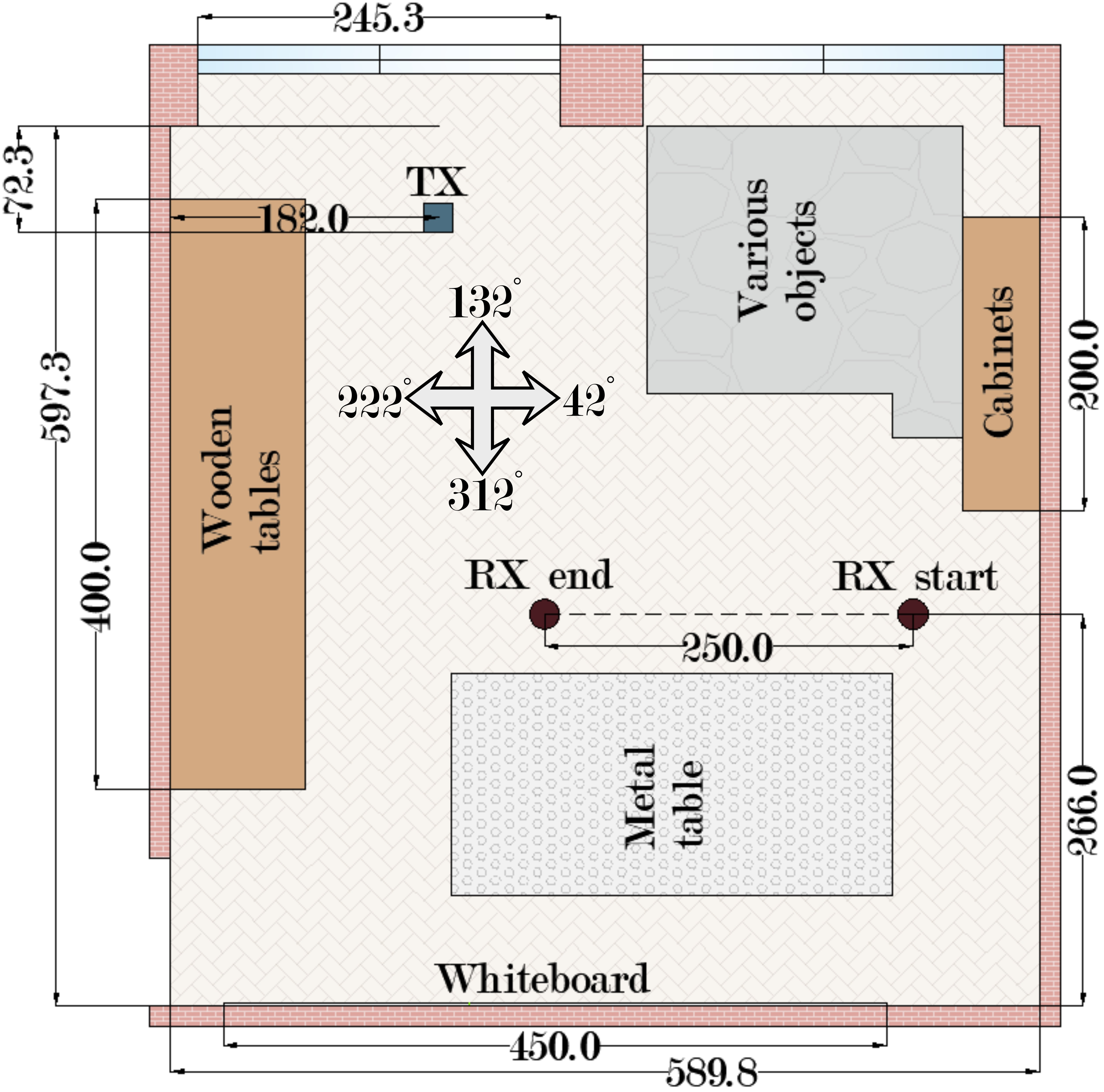}
         \caption{Layout (dimensions in cm)}
         \label{fig:measurementLayoutSequences}
     \end{subfigure}
     \begin{subfigure}{0.44\columnwidth}
         \centering
         \includegraphics[width=\textwidth]{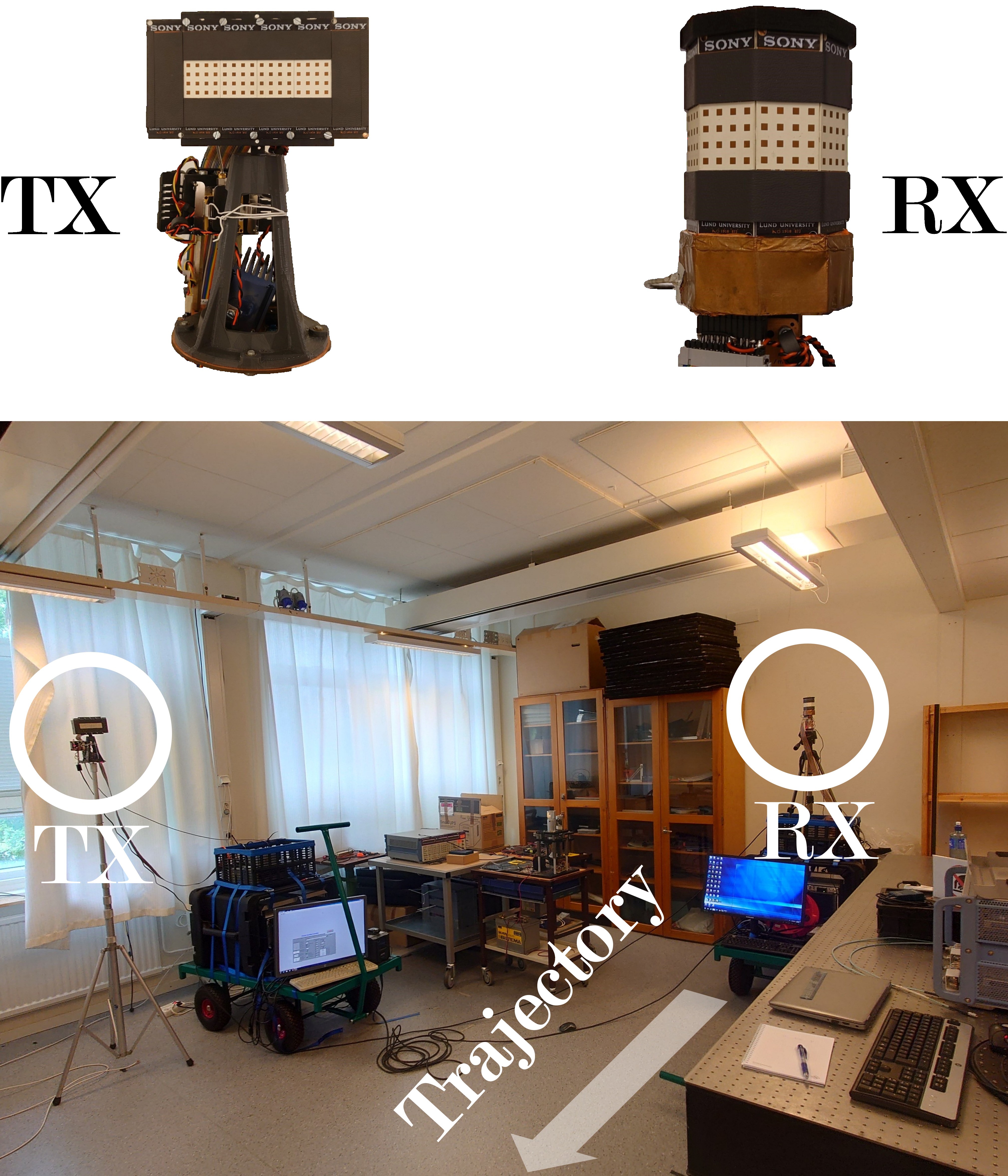}
         \caption{Photo of the setup}
         \label{fig:measurementPhotoSequences}
     \end{subfigure}
     \caption{Measurement environment.} \label{fig:measurementEnvironment}
\end{figure}

The performance of the Fourier-Fisher sequence design proposed in this paper was verified by double-directional polarimetric measurements with the Lund University mmWave channel sounder \cite{cai2024switched}. Fig. \ref{fig:measurementEnvironment} shows a sketch and a photo of the measurement setup, with Fig. \ref{fig:measurementPhotoSequences} also showing a close-up of the uniform planar array (UPA) and octagonal dual-polarized arrays, respectively deployed at the TX and RX sides, on top of the setup photo. Table \ref{tab:measurementParametersSequences} collects the most representative measurement parameters, together with the sequences under test. Each sequence dictated the activation order of the antennas at both TX and RX sides when measuring along a fixed track with the channel sounder. The RX antenna array was static for the first and last 5 seconds of the measurement, otherwise moving along the track with a constant speed. The measured channel impulse responses were then processed using a space-alternating generalized expectation-maximization (SAGE) implementation, where the MPCs were estimated for further analysis.



\begin{table}
\begin{center}
	\caption{Measurement setup parameters.}
	\label{tab:measurementParametersSequences}
	\begin{tabular}{|c|c|}
		
		\hline		
		\textbf{Channel sounder}	& Lund University \\&mmWave channel sounder \cite{cai2023enhanced}\\\hline
        \textbf{Carrier frequency}	& $28$\,GHz\\\hline
        \textbf{TX switching rate} & $4.8$\,ms\\\hline
        \textbf{RX switching rate} & $18.8$\,$\mu$s\\\hline
		\textbf{TX radio head} & $64$ dual-polarized UPA\\\hline
        \textbf{RX radio head} & $128$ dual-polarized octagonal array\\\hline
        \textbf{Track length} & $2.5$\,m\\\hline
        \textbf{Speed} & $10$\,cm/s\\\hline
        \textbf{Number of snapshots} & $35$\\\hline
        \textbf{Sequence 1} & Trivial\\\hline
        \textbf{Sequence 2} & Ambiguity \cite{wang2019channel}, $\mathcal{P} = 6$\\\hline
        \textbf{Sequence 3} & FF with realistic radiation patterns, \\&confidence level 99\%, \\&error margin 1\%\\\hline
	
	\end{tabular}
\end{center}
\end{table}

The switching sequences were generated using the Kronecker product approach explained in Sect.\,\ref{ss:kroneckerSwitching}, i.e., $\boldsymbol{\eta} = \boldsymbol{\eta}_{\text{T}} \otimes \boldsymbol{\eta}_{\text{R}}$. Therefore, the switching sequences can be optimized for TX and RX, respectively, with two independent Fourier optimization problems and two independent Fisher optimization problems. The Fisher cost functions $J_{\text{T}} \left( \boldsymbol{\eta}_{\text{T}} \right)$ and $J_{\text{R}} \left( \boldsymbol{\eta}_{\text{R}} \right)$ are developed from (\ref{eq:minCRLBGeneral}) as follows
\begin{equation}
\begin{aligned}
    J_{\text{T}}(\boldsymbol{\eta}_{\text{T}}) &= f_1\left( [\mathbf{F}]_{\nu\varphi_{\text{T}}} \right) + f_2\left( [\mathbf{F}]_{\nu\vartheta_{\text{T}}} \right)\\
    &= \max_{\varphi_{\text{T}}, \vartheta_{\text{T}}} \, \left| \Re \left\{ \left[ \mathbf{b}_{\text{T}} \odot \boldsymbol{\eta}_{\text{T}} \right]^{\hermitian} \mathbf{G}_{\text{T}} \left( \left[ \boldsymbol{\beta}_{\varphi_{\text{T}}} \odot \boldsymbol{\alpha}_{\varphi_{\text{T}}} \right] \otimes \boldsymbol{\beta}_{\vartheta_{\text{T}}} \right) \right\} \right|\\
    &+ \max_{\varphi_{\text{T}}, \vartheta_{\text{T}}} \, \left| \Re \left\{ \left[ \mathbf{b}_{\text{T}} \odot \boldsymbol{\eta}_{\text{T}} \right]^{\hermitian} \mathbf{G}_{\text{T}} \left( \boldsymbol{\beta}_{\varphi_{\text{T}}} \otimes \left[ \boldsymbol{\beta}_{\vartheta_{\text{T}}} \odot \boldsymbol{\alpha}_{\vartheta_{\text{T}}} \right] \right) \right\} \right|\\
\end{aligned}
\end{equation}
\begin{equation}
\begin{aligned}
    J_{\text{R}}(\boldsymbol{\eta}_{\text{R}}) &= f_3\left( [\mathbf{F}]_{\nu\varphi_{\text{R}}} \right) + f_4\left( [\mathbf{F}]_{\nu\vartheta_{\text{R}}} \right)\\
    &= \max_{\varphi_{\text{R}}, \vartheta_{\text{R}}} \, \left| \Re \left\{ \left[ \mathbf{b}_{\text{R}} \odot \boldsymbol{\eta}_{\text{R}} \right]^{\hermitian} \mathbf{G}_{\text{R}} \left( \left[ \boldsymbol{\beta}_{\varphi_{\text{R}}} \odot \boldsymbol{\alpha}_{\varphi_{\text{R}}} \right] \otimes \boldsymbol{\beta}_{\vartheta_{\text{R}}} \right) \right\} \right|\\
    &+ \max_{\varphi_{\text{R}}, \vartheta_{\text{R}}} \, \left| \Re \left\{ \left[ \mathbf{b}_{\text{R}} \odot \boldsymbol{\eta}_{\text{R}} \right]^{\hermitian} \mathbf{G}_{\text{R}} \left( \boldsymbol{\beta}_{\varphi_{\text{R}}} \otimes \left[ \boldsymbol{\beta}_{\vartheta_{\text{R}}} \odot \boldsymbol{\alpha}_{\vartheta_{\text{R}}} \right] \right) \right\} \right|\\
\end{aligned}
\end{equation}
where $\mathbf{b}_{\text{T}/\text{R}} = \mathbf{G}_{\text{T}/\text{R}} \cdot \left( \boldsymbol{\beta}_{\varphi_{\text{T}/\text{R}}} \otimes \boldsymbol{\beta}_{\vartheta_{\text{T}/\text{R}}} \right)$, $\boldsymbol{\beta}_{\varphi_{\text{T}/\text{R}}}$ is dependent on $\varphi_{\text{T}/\text{R}}$ and $\boldsymbol{\beta}_{\vartheta_{\text{T}/\text{R}}}$ is dependent on $\vartheta_{\text{T}/\text{R}}$.

\begin{figure}[t]
     \centering
     \begin{subfigure}{\columnwidth}
         \centering
         \includegraphics[width=\textwidth]{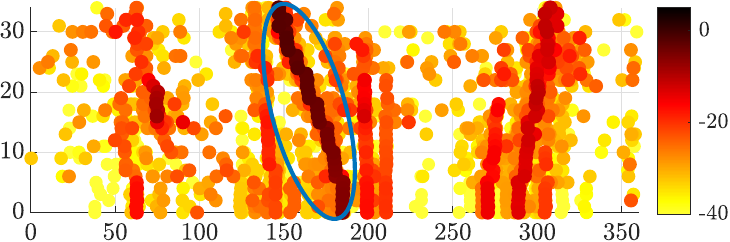}
         \caption{Trivial sequence}
         \label{fig:aoaTrivial}
     \end{subfigure}
     \begin{subfigure}{\columnwidth}
         \centering
         \includegraphics[width=\textwidth]{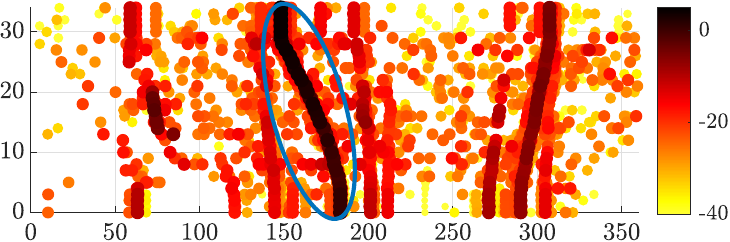}
         \caption{Ambiguity-based sequence}
         \label{fig:aoaAmbiguity}
     \end{subfigure}
     \begin{subfigure}{\columnwidth}
         \centering
         \includegraphics[width=\textwidth]{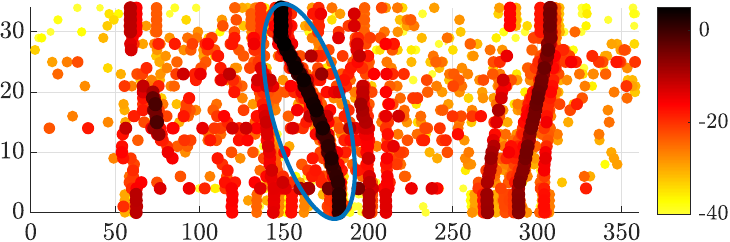}
         \caption{FF realistic sequence}
         \label{fig:aoaFFRealistic}
     \end{subfigure}
     \caption{The estimation results of MPC parameters using SAGE. X-axis: AOA [deg]; Y-axis: Snapshot index; Color axis: Power [dB].} \label{fig:mpcEstimatesSequences}
\end{figure}

Fig. \ref{fig:mpcEstimatesSequences} shows the MPC estimates for each of the switching sequences under test, represented across the AOA and position indices within the 2.5\,m trajectory. The evolution of the MPC points observed in the plots is due to the RX movement. It is clear from Fig. \ref{fig:aoaTrivial} that the performance of trivial switching is significantly affected by ambiguity, shown as discontinuities in the AOA estimation. Since multiple combinations of channel parameters can present a solution to an ambiguous estimation context, there are jumps in the estimates' evolution whenever one of such combinations attains the highest likelihood. For example, for the line of sight (LOS) path with AOA between 150-220\degrees, the piecewise AOA evolution when using trivial switching sequences does not match the reality of constant linear movement on the RX side. Figs. \ref{fig:aoaAmbiguity} and \ref{fig:aoaFFRealistic} show a better match for the mentioned evolution. The same goes for the rest of the MPC clusters, where evolution occurs in a smoother way when using either ambiguity or FF sequences, as expected with a constant speed in the measurement scenario. Furthermore, there are no noticeable differences in the estimation results found using either ambiguity or FF sequences. These observations are consistent with the results discussed in Sect.\,\ref{sec:simulationResults}, and show the good performance of the FF sequences in a real room. The plots in Fig.\,\ref{fig:mpcEstimatesSequences} also show blurred trajectories that are more evident for MPC estimates with lower power levels. One reason is that the low SNR levels of these paths lead to large estimation errors. Another reason is that the inherently imperfect system calibration can cause model mismatch, resulting in residual estimation errors perceived as ``ghost'' paths.

\section{Conclusions} \label{sec:conclusions}

This paper elaborated on multiple techniques to optimize switching sequences that improve channel sounding aided by theory, realistic simulations and measurements. A polarimetric spatio-ambiguity function was first introduced, and its computational complexity was analyzed to be cubic. The complexity can be reduced to linear when a high antenna cross-polarization ratio is assumed. The computation time can be further reduced by using a Kronecker-based switching structure in the switched arrays.

Since switching sequence design based on the ambiguity function is unfeasible for ultra-massive MIMO arrays, a novel design method with equivalent performance and significantly lower computation time was developed. The method was denoted ``Fourier-Fisher'', attributed to the two theoretical components involved in the process, namely Fourier transforms and Fisher information. The method first reduces the ambiguity side lobes by picking the switching sequence with the lowest median in its Fourier spectrum. The picked sequence is then refined via minimization of the off-diagonal elements of the Fisher information matrix associated with the parametric estimation problem. This iterative minimization features a configurable cost function and aims to reduce the width of the true estimation lobe, thus increasing precision. The method can be used in polarimetric channel sounders with high-XPR switched antenna arrays, as well as in wideband scenarios. Furthermore, enabling/disabling measured antenna radiation patterns as parameters in the cost function offers a key trade-off between performance and computation time.

Realistic simulations and measurements showed the ambiguity effect on the estimation of channel parameters and validated the competitive performance of the FF sequences, with a significantly reduced computation time from 30 up to 200 times lower than that of the ambiguity-based approach.


\begin{appendices}


\section{Proof of Theorem \ref{theorem:kroneckerOrthogonality}}
\label{app:proofKroneckerOrthogonality}

Given the matrices $\mathbf{A,B,C,D}$ of appropriate size, it is clear that $(\mathbf{A} \otimes \mathbf{B}) (\mathbf{C} \otimes \mathbf{D}) = (\mathbf{AC}) \otimes (\mathbf{BD})$, by the mixed-product property. Choosing $\mathbf{A} = \mathbf{u}^{\hermitian} \in \mathbb{C}^n$, $\mathbf{B} = \mathbf{w}^{\hermitian} \in \mathbb{C}^n$, $\mathbf{C} = \mathbf{v} \in \mathbb{C}^n$, $\mathbf{D} = \mathbf{w}$, it follows that
\begin{equation*}
\begin{aligned}
    \langle \mathbf{u} \otimes \mathbf{w} , \mathbf{v} \otimes \mathbf{w} \rangle &= (\mathbf{u} \otimes \mathbf{w})^{\hermitian} ( \mathbf{v} \otimes \mathbf{w}) = (\mathbf{u}^{\hermitian} \otimes \mathbf{w}^{\hermitian}) (\mathbf{v} \otimes \mathbf{w}) \\
    &= (\mathbf{u}^{\hermitian} \mathbf{v}) \otimes (\mathbf{w}^{\hermitian} \mathbf{w}) = (\mathbf{u}^{\hermitian} \mathbf{v}) \otimes c = 0,
\end{aligned}
\end{equation*}
where $c \in \mathbb{C}$, and the initial hypothesis was used. To prove that  $\langle \mathbf{w} \otimes \mathbf{u} , \mathbf{w} \otimes \mathbf{v} \rangle = 0$, choose $\mathbf{A} = \mathbf{w}^{\hermitian} \in \mathbb{C}^n$, $\mathbf{B} = \mathbf{u}^{\hermitian} \in \mathbb{C}^n$, $\mathbf{C} = \mathbf{w} \in \mathbb{C}^n$, $\mathbf{D} = \mathbf{v}$.

\section{Proof of Theorem \ref{theorem:hadamardOrthogonality}}
\label{app:proofHadamardOrthogonality}

\begin{equation*}
\begin{aligned}
    \langle \mathbf{u} \odot \mathbf{w} , \mathbf{v} \odot \mathbf{w} \rangle &= (\mathbf{u} \odot \mathbf{w})^{\hermitian} ( \mathbf{v} \odot \mathbf{w}) = \sum_{i = 0}^n \overline{(u_iw_i)}v_iw_i\\
    &= \sum_{i = 0}^n \bar{u_i}v_i||w_i||^2 = 0.
\end{aligned}
\end{equation*}
\begin{equation*}
\begin{aligned}
    \langle \mathbf{w} \odot \mathbf{u} , \mathbf{w} \odot \mathbf{v} \rangle &= (\mathbf{w} \odot \mathbf{u})^{\hermitian} ( \mathbf{w} \odot \mathbf{v}) = \sum_{i = 0}^n \overline{(w_iu_i)}w_iv_i\\
    &= \sum_{i = 0}^n \bar{u_i}v_i||w_i||^2 = 0.
\end{aligned}
\end{equation*}

\section{Proof of Lemma \ref{lemma:csPosDef}}
\label{app:proofCSPosDef}

Applying the Cauchy-Schwarz inequality,
\begin{equation*}
\begin{aligned}
    \sum_k \frac{b_k^2}{a_k^2} \sum_k a_k^2 b_k^2 &\geq \left( \sum_k \frac{b_k}{a_k} \cdot a_k b_k \right)^2 = \left( \sum_k b_k^2 \right)^2 = 1 \\
    \sum_k \frac{b_k^2}{a_k^2} &\geq \frac{1}{\sum_k a_k^2 b_k^2}.
\end{aligned}
\end{equation*}

\section{Proof of Theorem \ref{theorem:fisherCrossElements}}
\label{app:proofFisherCrossElements}

Using eigenvalue decomposition,
\begin{equation*}
\begin{aligned}
    \mathbf{A}^{-1} &= \left( \mathbf{U}\mathbf{\Lambda}\mathbf{U}^{\hermitian} \right)^{-1} = \left( \sum_j \lambda_j \cdot \mathbf{U}_j\mathbf{U}_j^{\hermitian} \right)^{-1} = \sum_j \frac{1}{\lambda_j} \cdot \mathbf{U}_j\mathbf{U}_j^{\hermitian} \\
    \left[ \mathbf{A}^{-1} \right]_{ii} &= \sum_j \frac{1}{\lambda_j} \cdot \mathbf{U}_{ij}\mathbf{U}_{ij}^{\hermitian} = \sum_j \frac{|\mathbf{U}_{ij}|^2}{\lambda_j},
\end{aligned}
\end{equation*}
where $\mathbf{U}$ is a unitary matrix, $\mathbf{U}_j$ is its j-th row, and $\mathbf{U}_{ij}$ is the element in its i-th column and j-th row. Moreover,
\begin{equation*}
\begin{aligned}
    \mathbf{B}^{-1} &= \left( \sum_k [\mathbf{A}]_{kk} \cdot e_k e_k^{\hermitian} \right)^{-1} = \sum_k \frac{1}{[\mathbf{A}]_{kk}} \cdot e_k e_k^{\hermitian} \\
    \left[ \mathbf{B}^{-1} \right]_{ii} &= \frac{1}{[\mathbf{A}]_{ii}} = \frac{1}{\sum_j \lambda_j|\mathbf{U}_{ij}|^2}.
\end{aligned}
\end{equation*}
Notice that $\sum_j |\mathbf{U}_{ij}|^2 = 1, \forall i$. We can then apply Lemma \ref{lemma:csPosDef} by letting $a_k = \sqrt{\lambda_j}$, $b_k = |\mathbf{U}_{ij}|$ and doing the correspondence between the $j$ and $k$ indices. For any index $i$,
\begin{equation*}
\begin{aligned}
    \left[ \mathbf{A}^{-1} \right]_{ii} = \sum_k \frac{|\mathbf{U}_{ij}|^2}{\lambda_j} &\geq \frac{1}{\sum_k \lambda_j |\mathbf{U}_{ij}|^2} = \left[ \mathbf{B}^{-1} \right]_{ii}.
\end{aligned}
\end{equation*}


\section{Derivation of the FIM entries using the EADF}
\label{app:fimEntriesGeneral}

The derivatives of the channel response with respect to the different directions of departure/arrival are
\begin{equation}
\begin{aligned} \label{eq:dvAODGeneral}
    \pdv{\mathbf{s} (\boldsymbol{\theta}_{\text{sp}})}{\varphi_{\text{T}}} = \left( \left[ \mathbf{G}_{\text{T}} \cdot \left( j \left[ \boldsymbol{\beta}_{\varphi_{\text{T}}} \odot \boldsymbol{\alpha}_{\varphi_{\text{T}}} \right] \otimes \boldsymbol{\beta}_{\vartheta_{\text{T}}} \right) \right] \otimes \mathbf{b}_{\text{R}}  \right) \odot \gamma\mathbf{a}_{\nu},
\end{aligned}
\end{equation}
\begin{equation}
\begin{aligned} \label{eq:dvEODGeneral}
    \pdv{\mathbf{s} (\boldsymbol{\theta}_{\text{sp}})}{\vartheta_{\text{T}}} = \left( \left[ \mathbf{G}_{\text{T}} \cdot \left( \boldsymbol{\beta}_{\varphi_{\text{T}}} \otimes j \left[ \boldsymbol{\beta}_{\vartheta_{\text{T}}} \odot \boldsymbol{\alpha}_{\vartheta_{\text{T}}} \right] \right) \right] \otimes \mathbf{b}_{\text{R}}  \right) \odot \gamma\mathbf{a}_{\nu},
\end{aligned}
\end{equation}
\begin{equation}
\begin{aligned} \label{eq:dvAOAGeneral}
    \pdv{\mathbf{s} (\boldsymbol{\theta}_{\text{sp}})}{\varphi_{\text{R}}} = \left( \mathbf{b}_{\text{T}} \otimes \left[ \mathbf{G}_{\text{R}} \cdot \left( j \left[ \boldsymbol{\beta}_{\varphi_{\text{R}}} \odot \boldsymbol{\alpha}_{\varphi_{\text{R}}} \right] \otimes \boldsymbol{\beta}_{\vartheta_{\text{R}}} \right) \right] \right) \odot \gamma\mathbf{a}_{\nu},
\end{aligned}
\end{equation}
\begin{equation}
\begin{aligned} \label{eq:dvEOAGeneral}
    \pdv{\mathbf{s} (\boldsymbol{\theta}_{\text{sp}})}{\vartheta_{\text{R}}} = \left( \mathbf{b}_{\text{T}} \otimes \left[ \mathbf{G}_{\text{R}} \cdot \left( \boldsymbol{\beta}_{\varphi_{\text{R}}} \otimes j \left[ \boldsymbol{\beta}_{\vartheta_{\text{R}}} \odot \boldsymbol{\alpha}_{\vartheta_{\text{R}}} \right] \right) \right] \right) \odot \gamma\mathbf{a}_{\nu}.
\end{aligned}
\end{equation}
The derivative of the channel response with respect to Doppler is
\begin{equation}
\begin{aligned} \label{eq:dvDopplerGeneral}
    \pdv{\mathbf{s} (\boldsymbol{\theta}_{\text{sp}})}{\nu} = \left( \mathbf{b}_{\text{T}} \otimes \mathbf{b}_{\text{R}} \right) \odot j2\pi\boldsymbol{\eta} \odot \gamma\mathbf{a}_{\nu}.
\end{aligned}
\end{equation}

Each of the FIM entries can be found by using (\ref{eq:fim}) and (\ref{eq:dvAODGeneral})-(\ref{eq:dvDopplerGeneral}). The diagonal entry of the FIM with respect to the AOD $\varphi_{\text{T}}$ can be derived as
\begingroup
\allowdisplaybreaks
\begin{alignat*}{3}
    [\mathbf{F}]_{\varphi_{\text{T}}\varphi_{\text{T}}} &= \frac{2}{\sigma^2} &&\Re \left\{ \left[ \pdv{\mathbf{s} (\boldsymbol{\theta}_{\text{sp}})}{\varphi_{\text{T}}} \right]^{\hermitian} \cdot \pdv{\mathbf{s} (\boldsymbol{\theta}_{\text{sp}})}{\varphi_{\text{T}}} \right\}\\
    &= \frac{2}{\sigma^2} &&\Re \left\{ \left[ \left( \left[ \mathbf{G}_{\text{T}} \cdot \left( j \left[ \boldsymbol{\beta}_{\varphi_{\text{T}}} \odot \boldsymbol{\alpha}_{\varphi_{\text{T}}} \right] \otimes \boldsymbol{\beta}_{\vartheta_{\text{T}}} \right) \right] \otimes \mathbf{b}_{\text{R}}  \right) \right]^{\hermitian} \right.\\
    & &&\left. \cdot \left[ \left( \left[ \mathbf{G}_{\text{T}} \cdot \left( j \left[ \boldsymbol{\beta}_{\varphi_{\text{T}}} \odot \boldsymbol{\alpha}_{\varphi_{\text{T}}} \right] \otimes \boldsymbol{\beta}_{\vartheta_{\text{T}}} \right) \right] \otimes \mathbf{b}_{\text{R}} \right) \right] \right\}\\
    & && \cdot ||\gamma\mathbf{a}_{\nu}||^2\\
    &= \frac{2r^2}{\sigma^2} &&\Re \left\{ \left[ \left( \left[ \mathbf{G}_{\text{T}} \cdot \left( j \left[ \boldsymbol{\beta}_{\varphi_{\text{T}}} \odot \boldsymbol{\alpha}_{\varphi_{\text{T}}} \right] \otimes \boldsymbol{\beta}_{\vartheta_{\text{T}}} \right) \right] \right) \right]^{\hermitian} \right.\\
    & &&\left. \cdot \left[ \left( \left[ \mathbf{G}_{\text{T}} \cdot \left( j \left[ \boldsymbol{\beta}_{\varphi_{\text{T}}} \odot \boldsymbol{\alpha}_{\varphi_{\text{T}}} \right] \otimes \boldsymbol{\beta}_{\vartheta_{\text{T}}} \right) \right] \right) \right] \right\}\\
    & && \otimes ( \mathbf{b}_{\text{R}}^{\hermitian} \cdot \mathbf{b}_{\text{R}} )\\
    &= \frac{2r^2}{\sigma^2} &&\Re \left\{ \left( \left[ \boldsymbol{\beta}_{\varphi_{\text{T}}}^{\hermitian} \odot \boldsymbol{\alpha}_{\varphi_{\text{T}}}^{\hermitian} \right] \otimes \boldsymbol{\beta}_{\vartheta_{\text{T}}}^{\hermitian} \right) \cdot \mathbf{G}_{\text{T}}^{\hermitian} \right.\\
    & &&\left.\ \cdot \mathbf{G}_{\text{T}} \cdot \left( \left[ \boldsymbol{\beta}_{\varphi_{\text{T}}} \odot \boldsymbol{\alpha}_{\varphi_{\text{T}}} \right] \otimes \boldsymbol{\beta}_{\vartheta_{\text{T}}} \right) \right\}, \numberthis
\end{alignat*}%
where the properties of the Hadamard product and the mixed-product for the Kronecker product were used. For the rest of the diagonal entries related to the directions of arrival/departure, the procedure is the same. The diagonal entry of the FIM with respect to Doppler can be derived as
\endgroup
\begin{equation}
\begin{aligned}
    [\mathbf{F}]_{\nu\nu} &= \frac{2}{\sigma^2} \Re \left\{ \left[ \pdv{\mathbf{s} (\boldsymbol{\theta}_{\text{sp}})}{\nu} \right]^{\hermitian} \cdot \pdv{\mathbf{s} (\boldsymbol{\theta}_{\text{sp}})}{\nu} \right\}\\
    &= \frac{2}{\sigma^2} \Re \left\{ \left( \mathbf{b}_{\text{T}} \otimes \mathbf{b}_{\text{R}} \right)^{\hermitian} \cdot \left( \mathbf{b}_{\text{T}} \otimes \mathbf{b}_{\text{R}} \right) \right\} \cdot ||2\pi\boldsymbol{\eta}||^2 \cdot ||\gamma\mathbf{a}_{\nu}||^2\\
    &= 8 \cdot \left(\frac{r\pi ||\boldsymbol{\eta}||}{\sigma}\right)^2.
\end{aligned}
\end{equation}
The off-diagonal entry of the FIM for cross Doppler-AOD is
\begingroup
\allowdisplaybreaks
\begin{alignat*}{2}
    [\mathbf{F}]_{\nu\varphi_{\text{T}}} &= \frac{2}{\sigma^2} \Re \left\{ \left[ \pdv{\mathbf{s} (\boldsymbol{\theta}_{\text{sp}})}{\nu} \right]^{\hermitian} \cdot \pdv{\mathbf{s} (\boldsymbol{\theta}_{\text{sp}})}{\varphi_{\text{T}}} \right\}\\
    &= \frac{2}{\sigma^2} \Re \left\{ \left[ \left( \mathbf{b}_{\text{T}} \otimes \mathbf{b}_{\text{R}} \right) \odot j2\pi\boldsymbol{\eta} \right]^{\hermitian} \right.\\
    &\left. \cdot \left( \left[ \mathbf{G}_{\text{T}} \cdot \left( j \left[ \boldsymbol{\beta}_{\varphi_{\text{T}}} \odot \boldsymbol{\alpha}_{\varphi_{\text{T}}} \right] \otimes \boldsymbol{\beta}_{\vartheta_{\text{T}}} \right) \right] \otimes \mathbf{b}_{\text{R}}  \right) \right\} \cdot ||\gamma\mathbf{a}_{\nu}||^2\\
    &= \frac{4 \pi r^2}{\sigma^2} \Re \left\{ \left( \mathbf{b}_{\text{T}} \otimes \mathbf{b}_{\text{R}} \right)^{\hermitian} \cdot \text{diag}\{\boldsymbol{\eta}\} \right.\\
    &\left. \cdot \left( \left[ \mathbf{G}_{\text{T}} \cdot \left( \left[ \boldsymbol{\beta}_{\varphi_{\text{T}}} \odot \boldsymbol{\alpha}_{\varphi_{\text{T}}} \right] \otimes \boldsymbol{\beta}_{\vartheta_{\text{T}}} \right) \right] \otimes \mathbf{b}_{\text{R}}  \right) \right\}\\
    &= \frac{4 \pi r^2}{\sigma^2} \Re \left\{ \boldsymbol{\eta}^{\transpose} \cdot \left[ \overline{\left( \mathbf{b}_{\text{T}} \otimes \mathbf{b}_{\text{R}} \right)} \right. \right.\\
    &\left. \left. \odot \left( \left[ \mathbf{G}_{\text{T}} \cdot \left( \left[ \boldsymbol{\beta}_{\varphi_{\text{T}}} \odot \boldsymbol{\alpha}_{\varphi_{\text{T}}} \right] \otimes \boldsymbol{\beta}_{\vartheta_{\text{T}}} \right) \right] \otimes \mathbf{b}_{\text{R}}  \right) \right] \right\}\\
    &= \frac{4 \pi r^2}{\sigma^2} \Re \left\{ \boldsymbol{\eta}^{\transpose} \cdot \left[ \left( \overline{\mathbf{b}_{\text{T}}} \odot \left[ \mathbf{G}_{\text{T}} \cdot \left( \left[ \boldsymbol{\beta}_{\varphi_{\text{T}}} \odot \boldsymbol{\alpha}_{\varphi_{\text{T}}} \right] \otimes \boldsymbol{\beta}_{\vartheta_{\text{T}}} \right) \right] \right) \right. \right.\\
    &\left. \left. \otimes \left( \overline{\mathbf{b}_{\text{R}}} \odot \mathbf{b}_{\text{R}}  \right) \right] \right\}\\
    &= \frac{4 \pi r^2}{\sigma^2} \Re \left\{ \boldsymbol{\eta}^{\transpose} \cdot \left[ \left( \overline{\mathbf{b}_{\text{T}}} \odot \left[ \mathbf{G}_{\text{T}} \cdot \left( \left[ \boldsymbol{\beta}_{\varphi_{\text{T}}} \odot \boldsymbol{\alpha}_{\varphi_{\text{T}}} \right] \otimes \boldsymbol{\beta}_{\vartheta_{\text{T}}} \right) \right] \right) \right. \right.\\
    &\left. \left. \otimes \mathbf{1}_{\text{M}_{\text{R}}} \right] \right\}\\
    &= \frac{4\pi r^2}{\sigma^2} \Re \left\{ \left[ \left( \mathbf{b}_{\text{T}} \otimes \mathbf{1}_{\text{M}_{\text{R}}} \right) \odot \boldsymbol{\eta} \right]^{\hermitian} \right.\\
    &\left. \cdot \left( \left[ \mathbf{G}_{\text{T}} \cdot \left( \left[ \boldsymbol{\beta}_{\varphi_{\text{T}}} \odot \boldsymbol{\alpha}_{\varphi_{\text{T}}} \right] \otimes \boldsymbol{\beta}_{\vartheta_{\text{T}}} \right) \right] \otimes \mathbf{1}_{\text{M}_{\text{R}}} \right) \right\}. \numberthis \label{eq:fimDopplerAOD}
\end{alignat*}
\endgroup
The remaining off-diagonal elements of cross Doppler-angle follow a similar derivation style. The off-diagonal entry of the FIM for cross AOD-AOA is
\begin{equation}
\begin{aligned}
    [\mathbf{F}]_{\varphi_{\text{T}}\varphi_{\text{R}}} &= \frac{2}{\sigma^2} \Re \left\{ \left[ \pdv{\mathbf{s} (\boldsymbol{\theta}_{\text{sp}})}{\varphi_{\text{T}}} \right]^{\hermitian} \cdot \pdv{\mathbf{s} (\boldsymbol{\theta}_{\text{sp}})}{\varphi_{\text{R}}} \right\}\\
    &= \frac{2}{\sigma^2} \Re \left\{ \left( \left[ \mathbf{G}_{\text{T}} \cdot \left( j \left[ \boldsymbol{\beta}_{\varphi_{\text{T}}} \odot \boldsymbol{\alpha}_{\varphi_{\text{T}}} \right] \otimes \boldsymbol{\beta}_{\vartheta_{\text{T}}} \right) \right] \otimes \mathbf{b}_{\text{R}}  \right)^{\hermitian} \right.\\
    &\left. \cdot \left( \mathbf{b}_{\text{T}} \otimes \left[ \mathbf{G}_{\text{R}} \cdot \left( j \left[ \boldsymbol{\beta}_{\varphi_{\text{R}}} \odot \boldsymbol{\alpha}_{\varphi_{\text{R}}} \right] \otimes \boldsymbol{\beta}_{\vartheta_{\text{R}}} \right) \right] \right) \right\} \cdot ||\gamma\mathbf{a}_{\nu}||^2\\
    &= \frac{2r^2}{\sigma^2} \Re \left\{ \left( \left[ \mathbf{G}_{\text{T}} \cdot \left( j \left[ \boldsymbol{\beta}_{\varphi_{\text{T}}} \odot \boldsymbol{\alpha}_{\varphi_{\text{T}}} \right] \otimes \boldsymbol{\beta}_{\vartheta_{\text{T}}} \right) \right] \otimes \mathbf{b}_{\text{R}}  \right)^{\hermitian} \right.\\
    &\left. \cdot \left( \mathbf{b}_{\text{T}} \otimes \left[ \mathbf{G}_{\text{R}} \cdot \left( j \left[ \boldsymbol{\beta}_{\varphi_{\text{R}}} \odot \boldsymbol{\alpha}_{\varphi_{\text{R}}} \right] \otimes \boldsymbol{\beta}_{\vartheta_{\text{R}}} \right) \right] \right) \right\}.
\end{aligned}
\end{equation}
This expression cannot be further reduced if no assumptions can be taken on the EADF matrices $\mathbf{G}_{\text{T}}$ and $\mathbf{G}_{\text{R}}$. The same goes for all off-diagonal elements of cross angle-angle.

\end{appendices}
\vspace{-1.8mm}

\bibliographystyle{IEEEtran}
\bibliography{ref}

\end{document}